\documentclass[dvips]{acta}
\usepackage{supertabular,lscape,epsfig}
\usepackage{amssymb}
\usepackage{amsmath}
\usepackage{float}
\mathchardef\mhyphen="2D
\usepackage{multirow}
\usepackage{placeins}%\Floatbarier
\DeclareSymbolFont{ppa}{OT1}{ppl}{m}{it}
\DeclareMathSymbol{\vv}{\mathalpha}{ppa}{'166}

\SetPages{213}{236}

\SetVol{68}{2018}

\renewcommand{\doi}{doi: 10.32023/0001-5237/68.3.3}

\begin{document}
\newcommand\pvalue{\mathop{p\mhyphen {\rm value}}}
%Zwarte naglowki, jeden wiersz, appendix
\newcommand{\TabApp}[2]{\begin{center}\parbox[t]{#1}{\centerline{
  {\bf Appendix}}
  \vskip2mm
  \centerline{\small {\spaceskip 2pt plus 1pt minus 1pt T a b l e}
  \refstepcounter{table}\thetable}
  \vskip2mm
  \centerline{\footnotesize #2}}
  \vskip3mm
\end{center}}

%Zwarte naglowki, jeden wiersz
\newcommand{\TabCapp}[2]{\begin{center}\parbox[t]{#1}{\centerline{
  \small {\spaceskip 2pt plus 1pt minus 1pt T a b l e}
  \refstepcounter{table}\thetable}
  \vskip2mm
  \centerline{\footnotesize #2}}
  \vskip3mm
\end{center}}

%Zwarte naglowki, dwa wiersze
\newcommand{\TTabCap}[3]{\begin{center}\parbox[t]{#1}{\centerline{
  \small {\spaceskip 2pt plus 1pt minus 1pt T a b l e}
  \refstepcounter{table}\thetable}
  \vskip2mm
  \centerline{\footnotesize #2}
  \centerline{\footnotesize #3}}
  \vskip1mm
\end{center}}

%Zwarte naglowki, jeden wiersz, appendix
\newcommand{\MakeTableApp}[4]{\begin{table}[p]\TabApp{#2}{#3}
  \begin{center} \TableFont \begin{tabular}{#1} #4 
  \end{tabular}\end{center}\end{table}}

%Zwarte naglowki, jeden wiersz
\newcommand{\MakeTableSepp}[4]{\begin{table}[p]\TabCapp{#2}{#3}
  \begin{center} \TableFont \begin{tabular}{#1} #4 
  \end{tabular}\end{center}\end{table}}

%Zwarte naglowki, jeden wiersz
\newcommand{\MakeTableee}[4]{\begin{table}[htb]\TabCapp{#2}{#3}
  \begin{center} \TableFont \begin{tabular}{#1} #4
  \end{tabular}\end{center}\end{table}}

%Zwarte naglowki, dwa wiersze
\newcommand{\MakeTablee}[5]{\begin{table}[htb]\TTabCap{#2}{#3}{#4}
  \begin{center} \TableFont \begin{tabular}{#1} #5 
  \end{tabular}\end{center}\end{table}}

%Tabela w okre¶lonym miejscu
\newcommand{\MakeTableH}[4]{\begin{table}[H]\TabCap{#2}{#3}
  \begin{center} \TableFont \begin{tabular}{#1} #4 
  \end{tabular}\end{center}\end{table}}

%Tabela w okre¶lonym miejscu, zwatre nag³ówki, jeden wiersz
\newcommand{\MakeTableHH}[4]{\begin{table}[H]\TabCapp{#2}{#3}
  \begin{center} \TableFont \begin{tabular}{#1} #4 
  \end{tabular}\end{center}\end{table}}

%wyrównanie w tabeli - wzglêdem kropki dziesiêtnej r@.l
%{\it Acta Astronomica Archive}
%\parskip=0pt \itemsep=1mm \setlength{\itemsep}{0.4mm}\setlength{\parindent}{-1em} \setlength{\itemindent}{-1em} - po \begin{itemize} - wszystko
%FWHM, PSF, S/N - proste, 
%MgII, H$\alpha$
%rms, rhs, sd - kursywa
%{\sc DAOPhot}
%{\sc Fnpeaks}
%{\sf files}
%Galactic wszystko (bulge, center, plane, disk, coordinates, latitudes...)
%Cepheids
%type~ Cepheids, Population~II Cepheids
%a.u. => au (od AcA 3/2018)
%Polish National Science Centre
\newfont{\bb}{ptmbi8t at 12pt}
\newfont{\bbb}{cmbxti10}
\newfont{\bbbb}{cmbxti10 at 9pt}
\newcommand{\uprule}{\rule{0pt}{2.5ex}}
\newcommand{\douprule}{\rule[-2ex]{0pt}{4.5ex}}
\newcommand{\dorule}{\rule[-2ex]{0pt}{2ex}}
\begin{Titlepage}
\Title{Three-Dimensional Distributions of Type II Cepheids and Anomalous
Cepheids in the Magellanic Clouds.\\ 
Do these Stars Belong to the Old, Young or Intermediate-Age Population?}
\Author{P.~~I~w~a~n~e~k$^1$,~~
I.~~S~o~s~z~y~ñ~s~k~i$^1$,~~ 
D.~~S~k~o~w~r~o~n$^1$,~~ J.~~S~k~o~w~r~o~n$^1$,\\
P.~~M~r~ó~z$^1$,~~ 
S.~~K~o~z~³~o~w~s~k~i$^1$,~~ 
A.~~U~d~a~l~s~k~i$^1$,~~ 
M.\,K.~~S~z~y~m~a~ñ~s~k~i$^1$,\\
P.~~P~i~e~t~r~u~k~o~w~i~c~z$^1$,~~ 
R.~~P~o~l~e~s~k~i$^{2,1}$~~ 
and~~ A.~~J~a~c~y~s~z~y~n~-~D~o~b~r~z~e~n~i~e~c~k~a$^1$}
{$^1$Warsaw University Observatory, Al. Ujazdowskie 4, 00-478 Warsaw, Poland\\
e-mail: piwanek@astrouw.edu.pl\\
$^2$Department of Astronomy, Ohio State University, 140 W. 18th Ave., Columbus, OH 43210, USA}
\Received{September 9, 2018}
\end{Titlepage}

\Abstract{The nature of type II Cepheids and anomalous Cepheids is
  still not well known and their evolutionary channels leave many
  unanswered questions. We use complete collection of classical
  pulsating stars in the Magellanic Clouds discovered by the OGLE
  project, to compare their spatial distributions, which are one of
  the characteristic features directly related to the star formation
  history. In this analysis we use 9649 classical Cepheids, 262
  anomalous Cepheids, 338 type II Cepheids and 46\,443 RR~Lyr stars
  from both Magellanic Clouds. We compute three-dimensional
  Kolmogorov-Smirnov tests for every possible pair of type II and
  anomalous Cepheids with classical Cepheids, and RR~Lyr stars.
  We~confirm that BL~Her stars are as old as RR~Lyr variable stars --
  their spatial distributions are similar, and they create a vast halo
  around both galaxies. We discover that spatial distribution of W~Vir
  stars has attributes characteristic for both young and old stellar
  populations. Hence, it seems that these similarities are related to
  the concentration of these stars in the center of the Large
  Magellanic Cloud, and the lack of a vast halo. This leads to the
  conclusion that W~Vir variables could be a mixture of old and
  intermediate-age stars. Our analysis of the three-dimensional
  distributions of anomalous Cepheids shows that they differ
  significantly from classical Cepheids. Statistical tests of
  anomalous Cepheids distributions with RR~Lyr distributions do not
  give unambiguous results. We consider that these two distributions
  can be similar through the vast halos they create. This similarity
  would confirm anomalous Cepheids evolution scenario that assumes
  coalescence of a binary system.}{Stars: variables: Cepheids -- Stars:
  variables: RR Lyrae -- Magellanic Clouds}

\Section{Introduction}
Many years of research on the pulsating stars, including famous
discovery of the period--luminosity (PL) relation (Leavitt and
Pickering 1912), made classical pulsators the primary distance
indicators in the nearby Universe. Subsequent discoveries of pulsating
stars led to the distance scale revision done by Baade (1952), who
noted that there are two different groups of Cepheids which follow
different PL relations. Hence, the Cepheids were divided into
Population I (called classical Cepheids -- DCEPs) and Population II
(called type II Cepheids -- T2CEPs). Nowadays, T2CEPs are divided into
three subgroups depending on their pulsation period $P$: BL~Her
($P<4$~d), W~Vir ($4~{\rm d}\leq P<20$~d) and RV~Tau ($P>20$~d). The
boundaries in the pulsation periods for these stars are not strict,
and these groups partly overlap. Additionally, Soszyñski \etal (2008)
distinguished a fourth subgroup of T2CEPs, named peculiar W~Vir stars.
\vskip3pt
After over sixty years of study of T2CEPs their origin and evolution
channels are still not clear. The first evolution scenarios of these
stars were proposed by Gingold (1976, 1985). The most up-to-date
review of the T2CEPs properties has been done by Welch (2012). It is
believed that these objects are low-mass stars belonging to the halo
and old disk stellar populations. BL~Her stars evolve from blue
horizontal giant branch to asymptotic giant branch. During evolution
these stars become brighter and their radii increase, which is
associated with helium burning in the cores. BL~Her variables pass
through the instability strip at luminosities, which correspond to the
pulsation periods shorter than 4~d. RV~Tau variables are
post-asymptotic giant branch stars just before the outer envelope
expulsion, which is supposed to form a~planetary nebula, and a core
that becomes a white dwarf. These stars pass through the
high-luminosity extension of the Cepheid instability strip which
corresponds to pulsation periods longer than 20~d. The origin of W~Vir
stars is the most incomprehensible. It is believed that these
variables are asymptotic giant branch stars that have exhausted helium
in their cores, and they start burning helium in the shell. The helium
shell burning causes a gradual reduction of the energy supply from the
hydrogen shell, which in turn leads to the stoppage of energy
production in the hydrogen shell. The hydrogen shell burning may
re-switch due to heating arising from contraction. Switching on and
off of the shells burning may cause blue loops in the instability
strip. The loops are expected only in stars with small enough external
envelope, and they could be recurring (Cassisi and Salaris 2013), but
modern calculations are not able to reproduce this scenario. So far,
there are no evolutionary channels that would explain the origin of
W~Vir variable stars (Groenewegen and Jurkovic 2017a).
\vskip3pt
Soszyñski \etal (2008) found evidences that peculiar W~Vir stars have
companions. The evolution in a binary system provides conditions
conducive to the occurrence of pulsations. Up-to-date, 50\% of known
peculiar W~Vir stars from the Magellanic Clouds have clear signs of
binarity in the light curves.
\vskip3pt
In Fig.~1 we present distributions of T2CEPs in the sky. Positions of
BL~Her and W~Vir stars show a large scatter in both Magellanic
Clouds. RV~Tau stars seem to be more concentrated around the centers
of the Large Magellanic Cloud (LMC)\break
\begin{landscape}
\begin{figure}[htb]
\centerline{\includegraphics[width=9.7cm]{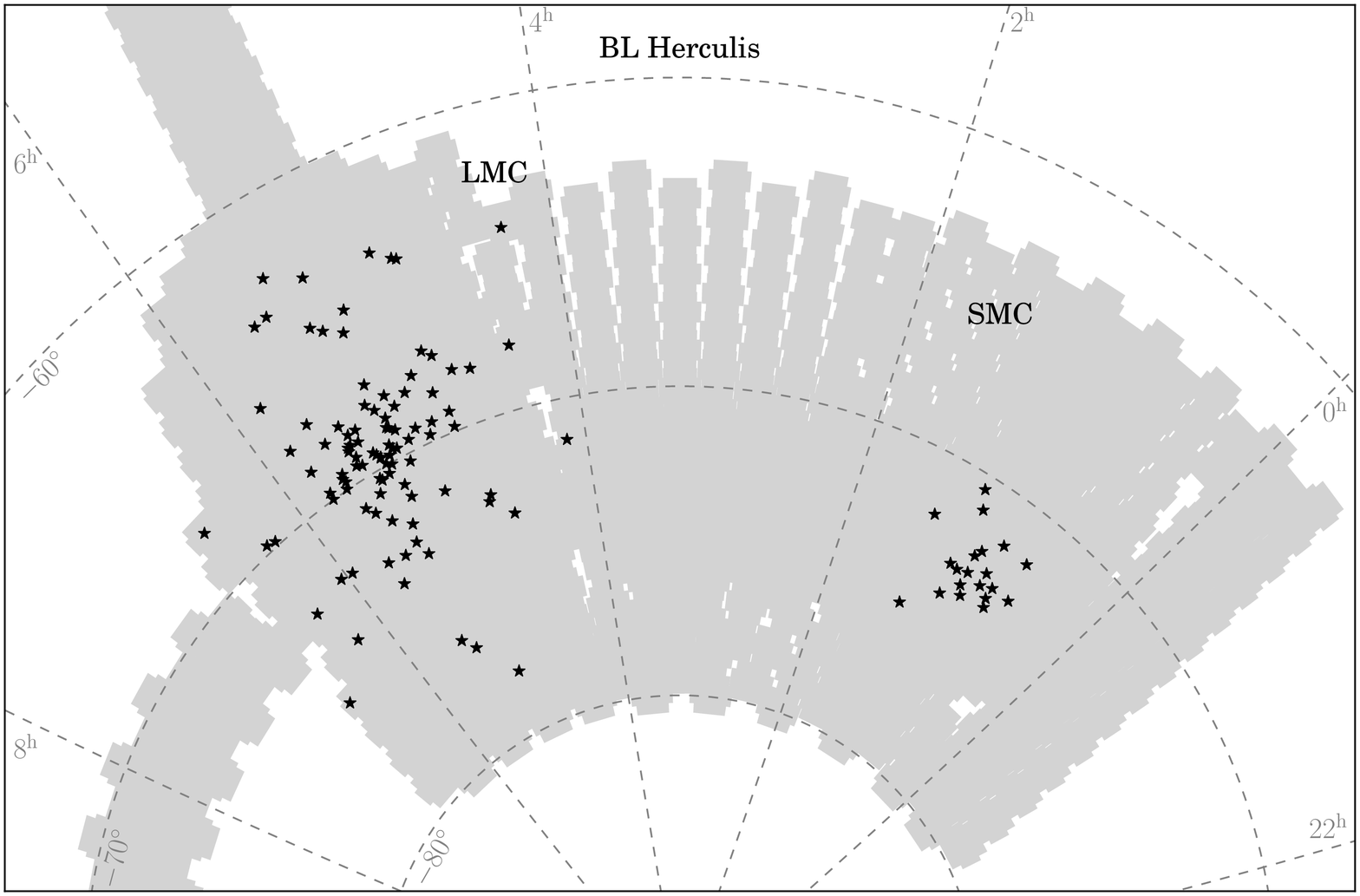} \includegraphics[width=9.7cm]{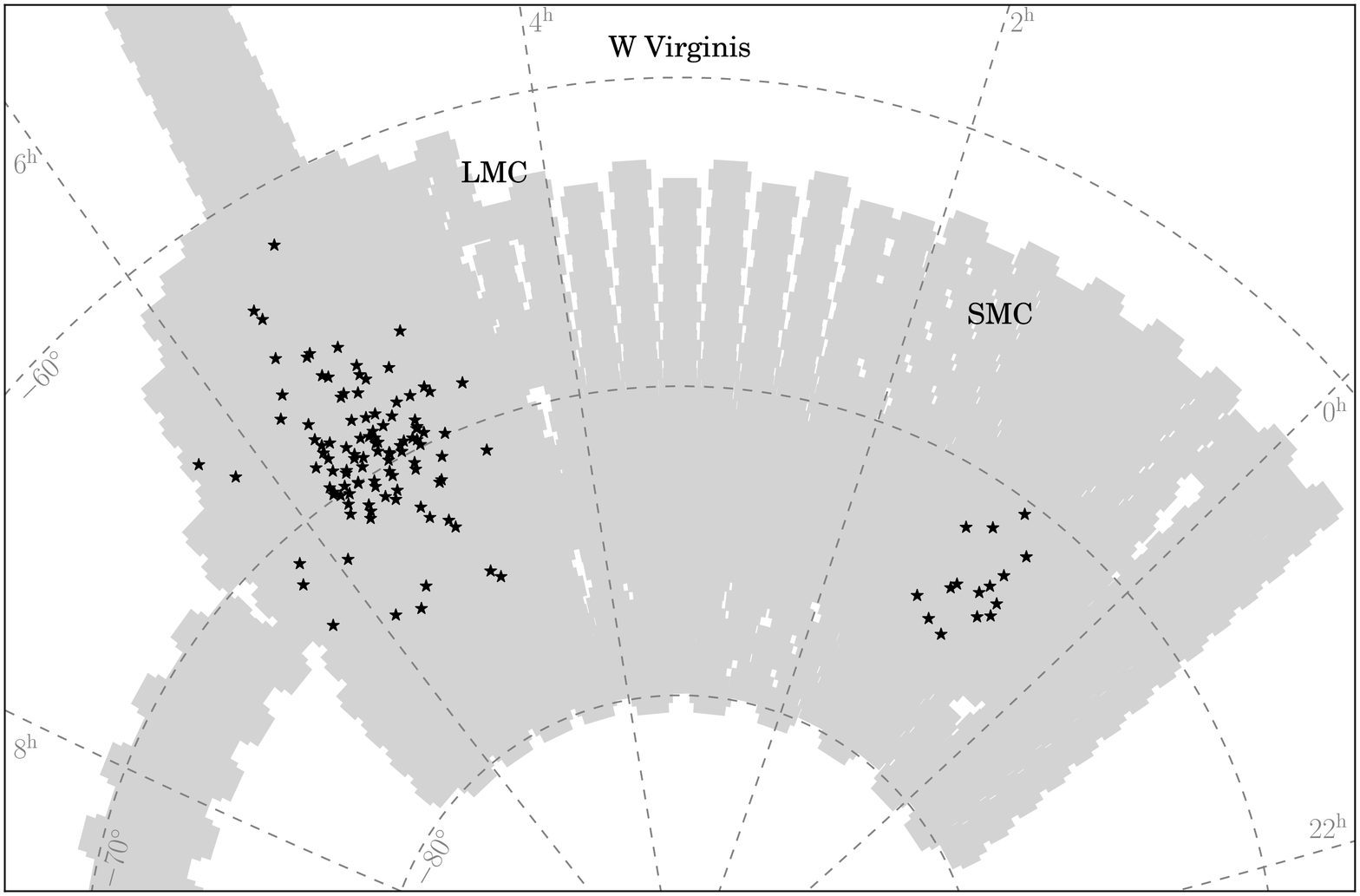}}
\centerline{\includegraphics[width=9.7cm]{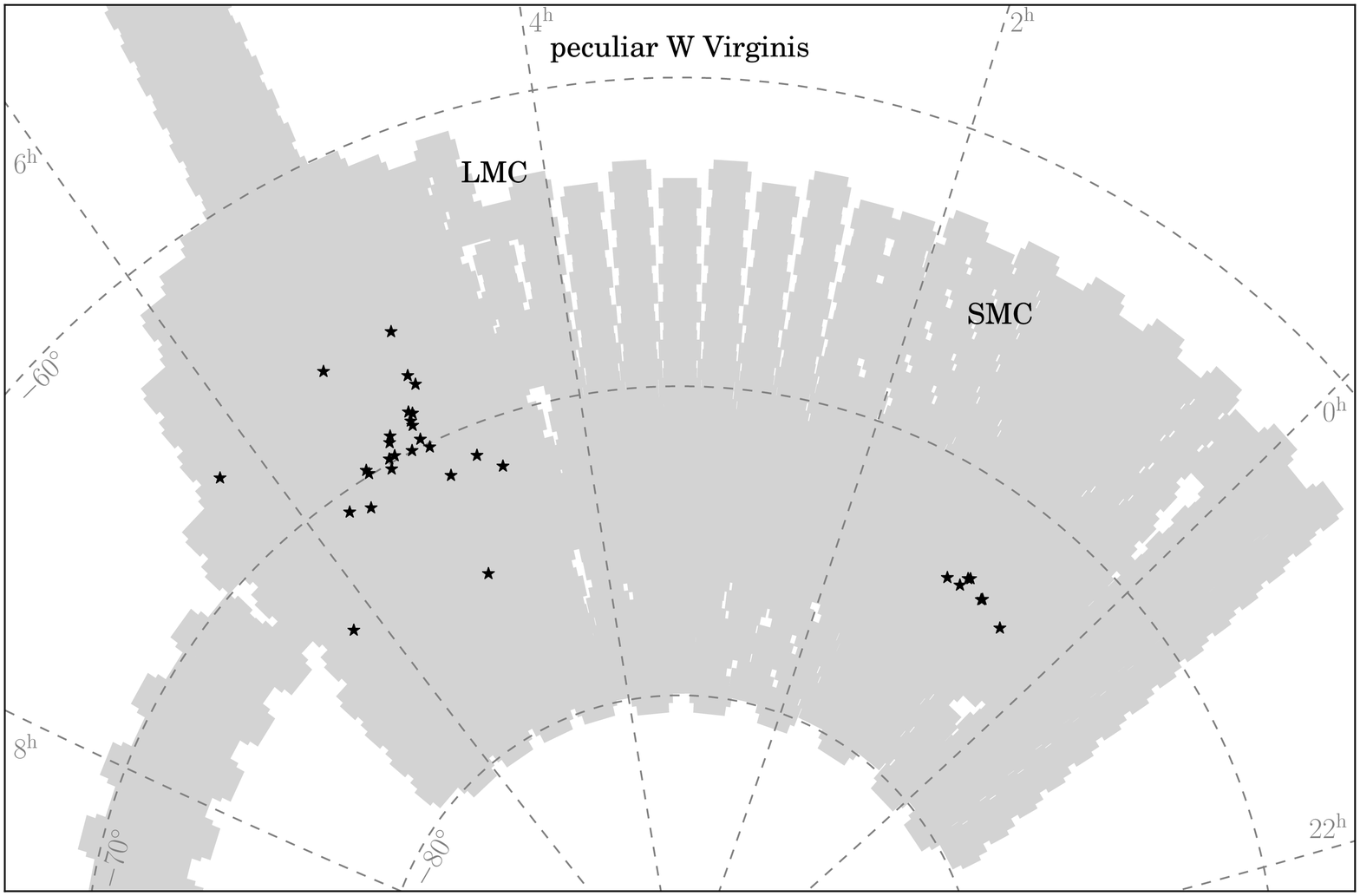} \includegraphics[width=9.7cm]{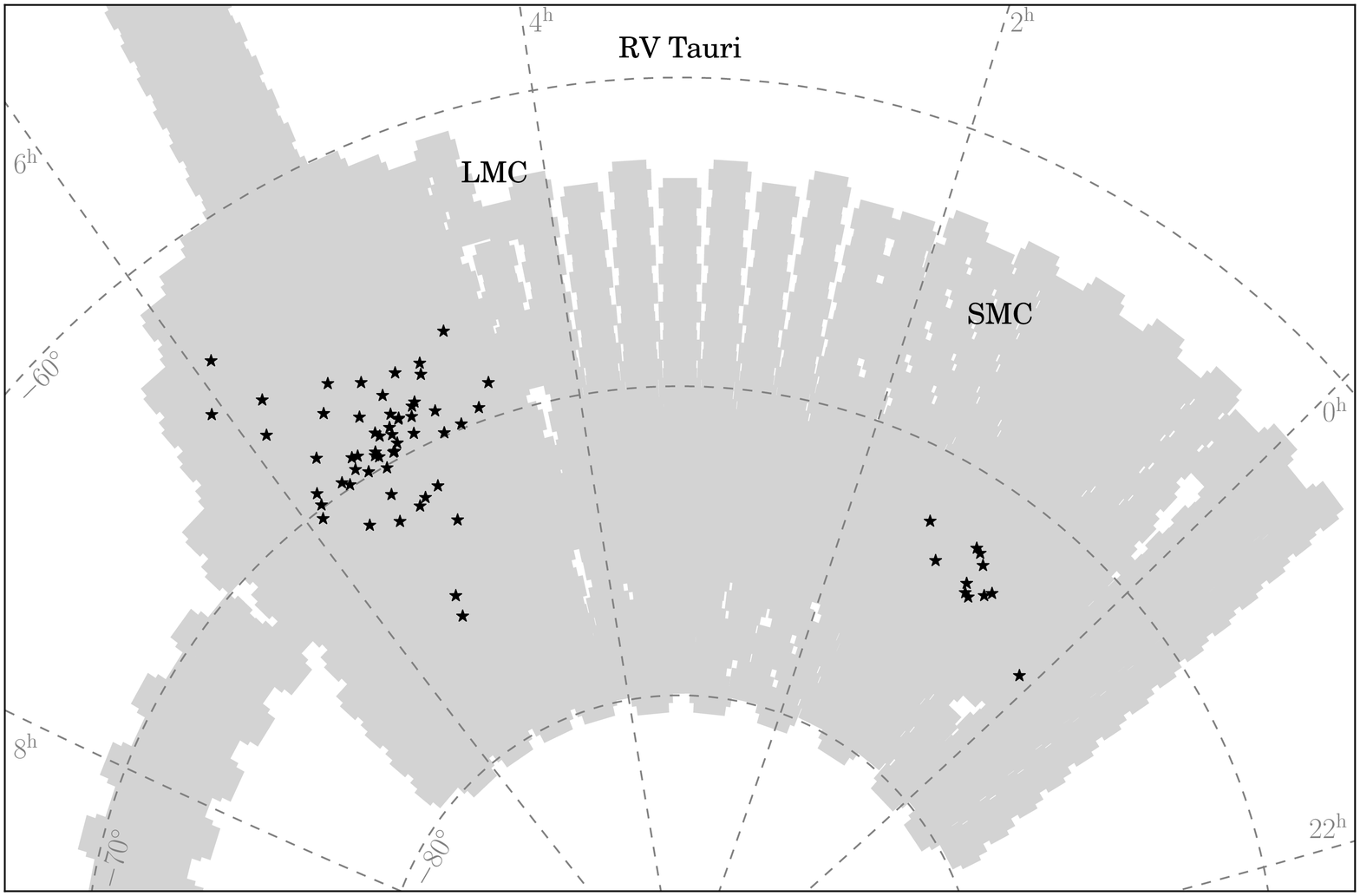}}
\FigCap{Distributions of T2CEPs in the sky. The gray area presents the OGLE-IV footprint.}
\end{figure}
\end{landscape}
\noindent 
and Small Magellanic Cloud (SMC). Peculiar W~Vir stars have slightly
different positions in the sky than other T2CEPs -- they are mostly
clumped around the bar of the LMC and they are located in the center
of the SMC. In addition, T2CEPs are found in globular clusters in the
LMC, with exception of peculiar W~Vir stars (Matsunaga \etal
2009). All these properties suggest that BL~Her, W~Vir and RV~Tau
variables very likely belong to the old population, while peculiar
W~Vir stars may be younger.

Observations of dwarf galaxies in the 50s and 60s of the 20th century
showed the existence of another group of pulsating stars, which
brightness variations did not match any of the known groups (Thackeray
1950). For this reason this group has been called anomalous Cepheids
-- ACEPs (Zinn and Searle 1976). The existence of this group of
pulsating stars can be explained in two ways: by evolution of single
intermediate-mass, metal-deficient star, which burns helium in the
core, or as the effect of coalescence of two old, low-mass stars which
evolved in the binary system. Fiorentino and Monelli (2012) found that
ACEPs distribution is different than DCEPs or RR~Lyr stars
distributions. They also suggested that observations of the LMC
outskirts could be helpful to solve the problem of ACEPs origin.

In Fig.~2, we present sky distribution of ACEPs in the LMC, and
SMC. It is clearly seen that these type of pulsating stars create a
vast halo around the Clouds. Moreover, there are a few objects in the
area between the Magellanic Clouds (called Magellanic
Bridge). Therefore, the distribution suggests that these stars belong to the
population as old as RR~Lyr variables or even older, or they could
reside in the Galactic foreground.
\begin{figure}[b]
\centerline{\includegraphics[width=9.7cm]{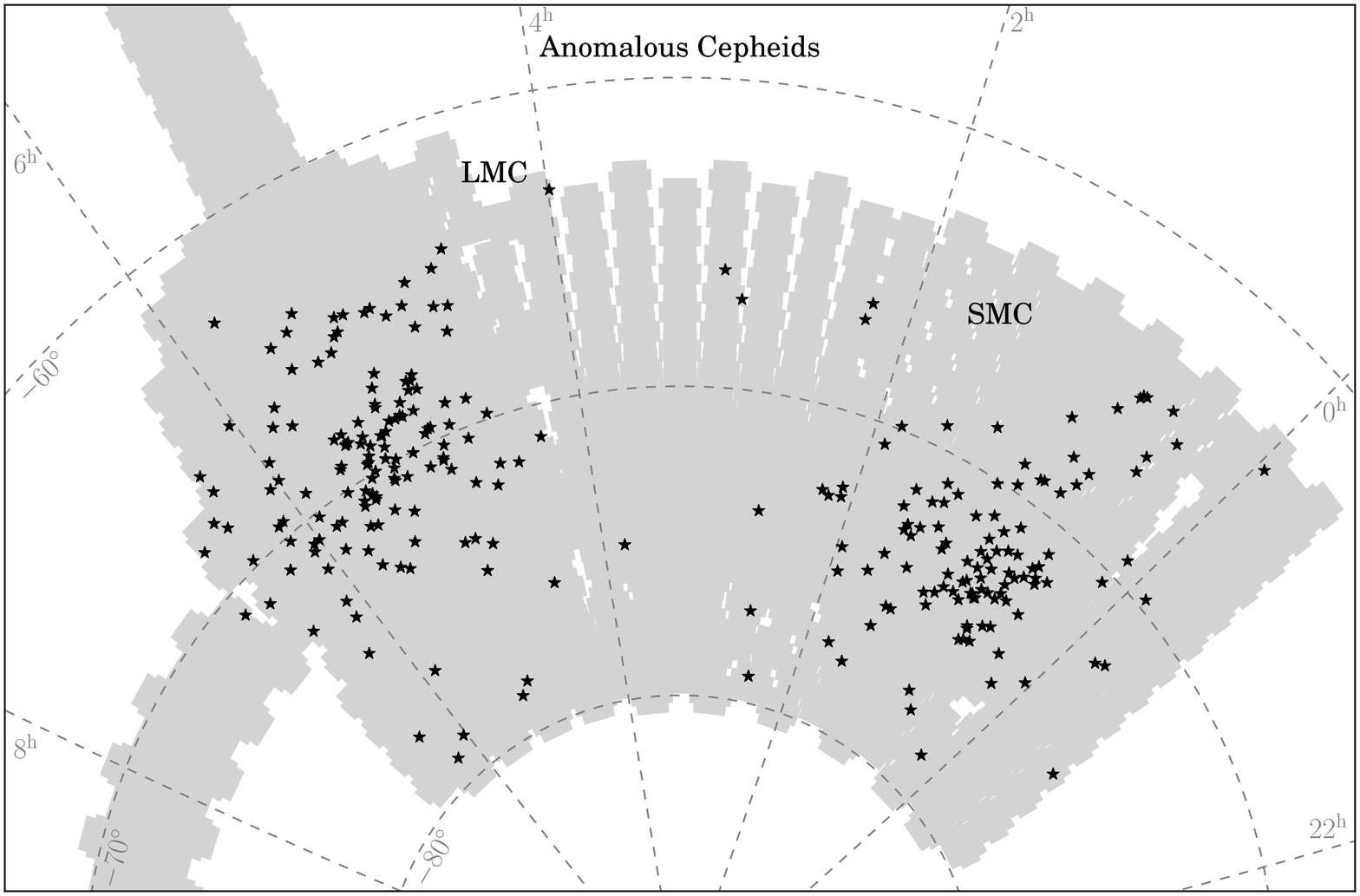}}
\FigCap{Distribution of ACEPs in the sky. The gray area presents the OGLE-IV footprint.}
\end{figure}

In this paper we use our complete collection of pulsating stars
discovered in the OGLE-IV data to compare three-dimensional
distributions of pulsators in the Magellanic Clouds. Our knowledge
about evolutionary status of DCEPs and RR~Lyr stars is more complete
compared to what we know about T2CEPs or ACEPs. We believe that the
comparison of the distributions of T2CEPs and ACEPs with other
classical pulsators distributions could shed light on the history and
the future of these stars. The spatial distribution of T2CEPs and
ACEPs may be a key to understanding their origin.

The paper is organized as follows. In Section~2, we discuss pulsating
star samples selection, which we use in our analysis. The distance
determination method used in this paper and transformation of coordinates
to the Cartesian space and Hammer equal-area projection are presented
in Section~3. Section~4 contains a detailed discussion of the method
of performing three-dimensional Kolmogorov-Smirnov statistical
test. In Section~5, we discuss results of our analysis with probable
evolution scenarios for each group of T2CEPs and ACEPs. We compare the
decision-making methods in Section 6. Finally, in Section~7 we
summarize our results.

\vspace*{7pt}
\Section{Sample Selection}
\vspace*{3pt}
The photometric data obtained by the Optical Gravitational Lensing
Experiment (OGLE) over 25 years of its activity has allowed to
increase the number of known classical pulsating stars by a large
factor. At this moment our collection contains 9649~DCEPs, 262~ACEPs,
338~T2CEPs and 46\,443 RR~Lyr stars in the Magellanic System
(Soszyñski \etal 2017), which is the most complete and the largest
list of classical pulsators in these galaxies. The entire collection
is available on-line {\it via} the OGLE FTP site:
\begin{center}
{\it ftp://ftp.astrouw.edu.pl/ogle/ogle4/OCVS/}
\end{center}
Details about the OGLE instrumentation, data reductions, calibrations,
sky coverage and observing cadence can be found in Udalski \etal
(2015).

In our study, we use stars with a single radial mode excited. It is
worth mentioning, however, that a recent research showed presence of
additional low-amplitude periodicities in stars classified as
single-mode pulsators. Netzel \etal (2015) showed that 27\% of the
first overtone RR~Lyr stars (RRc) have additional periodicities, which
can be interpreted as caused by non-radial modes (Dziembowski \etal
2016). Moskalik \etal (2015) analyzed observations of RRc stars from
the Kepler satellite and they found low-amplitude non-radial modes
excited in every considered star. The same phenomenon was observed in
the first overtone DCEPs (Soszyñski \etal 2015c, Smolec and {\'S}niegowska
2016, S\"uveges and Anderson 2018) and fundamental-mode pulsators (\eg
Smolec \etal 2016, Prudil \etal 2017). The additional periodicities
usually have low amplitudes and so do not influence the inferred
distances.

Another noteworthy aspect is an amplitude and phase modulations
present in some single-mode stars. The most pronounced effect of this
type with a high incidence rate is the Blazhko effect. This phenomenon
in noticeable in the fundamental-mode RR~Lyr stars (RRab -- Prudil and
Skarka 2017), and recently Netzel \etal (2018) showed that 5.6\% of RRc
stars from the OGLE sample show Blazhko modulations.

The OGLE collection of DCEPs in the Magellanic Clouds consist of 9649
stars (Soszyñski \etal 2015b, 2017). For our analysis we use DCEPs
which pulsate solely in the fundamental mode (F-mode -- 5229 objects
in total) or solely in the first overtone (1O -- 3568 objects in
total) as the most numerous samples. 2476 of the F-mode and 1775 of
the 1O DCEPs are located in the LMC, whereas 2753 F-mode and 1793 1O
DCEPs are located in the SMC.

The OGLE sample of pulsating stars consists of 46\,443 RR~Lyr
variables (So\-szyñski \etal 2016, 2017). We choose RR~Lyr stars
which pulsate solely in the fundamental mode (RRab -- 33\,297 objects)
or solely in the first overtone (RRc -- 10\,464 objects). 28\,192 RRab
stars are located in the LMC and the rest of the sample (5105 objects)
are located in the SMC. In the case of the RRc stars we have 9663
objects in the LMC and 801 stars in the SMC. Some of these stars may
belong to the Milky Way halo. In the LMC center blending and crowding
effects may influence the three-dimensional distribution of RR~Lyr
stars (Jacyszyn-Dobrzeniecka \etal 2017). To partially eliminate
blended RRab variables, we reject all objects for which peak-to-peak
{\it I}-band amplitudes $A_I< -5\cdot\log(P)-1$, where $P$ is the
pulsating period in the F-mode (the Bailey diagram, see Fig.~1 in
Jacyszyn-Dobrzeniecka \etal 2017). After this rejection we are left
with 26\,681 RRab stars in the LMC and 5018 objects in the SMC.

Our collection of T2CEPs in the Magellanic Clouds contains 338 stars
in total (285 stars in the LMC and 53 objects in the SMC,
Soszyñski \etal 2017, 2018). We subdivide T2CEPs by the pulsation
period. In the entire collection, we have 98 BL~Her stars in the LMC,
and 20 stars in the SMC, 106 W~Vir objects in the LMC, and 15 objects
in the SMC, and 55 RV~Tau stars in the LMC, and 11 such stars in the
SMC. The least numerous group is peculiar W~Vir stars. Our sample
contains 26 these objects in the LMC, and 7 objects in the SMC.

To date, the OGLE project has found 262 ACEPs in the Magellanic System
(Soszyñski \etal 2015a, 2017). As before, we choose for our
analysis stars which pulsate solely in the fundamental mode or solely
in the first overtone. 102 F-mode pulsators are located in the LMC,
and 78 are located in the SMC. In the case of the 1O ACEPs, we have 41
objects in the LMC, and the same number in the SMC.

\MakeTableee{cl|c|c|r}{12.5cm}{PL relations for each type of pulsating stars in 
the Magellanic Clouds analyzed in this paper}
{\hline
\douprule
Galaxy                & Type of stars             & $a$                & $b$                 & $N_{\rm fin}$\\ 
\hline
\uprule\multirow{8}{*}{LMC}  & F-mode Cepheids & $-3.319\pm0.008$ & $15.892\pm0.005$  & 2203 \\
                             & 1O Cepheids     & $-3.440\pm0.008$ & $15.398\pm0.003$  & 1594 \\ 
                             & RRab            & $-2.975\pm0.016$ & $17.167\pm0.004$  & 23265\\
                             & RRc             & $-3.109\pm0.026$ & $16.682\pm0.013$  & 8376 \\
		             & BL~Her          & $-2.683\pm0.091$ & $17.356\pm0.024$  & 79   \\                      
		             & W~Vir           & $-2.536\pm0.060$ & $17.378\pm0.062$  & 94   \\
                             & F-mode ACEPs    & $-2.957\pm0.118$ & $16.591\pm0.018$  & 94   \\
\dorule                      & 1O ACEPs        & $-3.298\pm0.200$ & $16.041\pm0.041$  & 39   \\
                      \hline
\uprule\multirow{8}{*}{SMC}  & F-mode Cepheids & $-3.448\pm0.011$ & $16.496\pm0.005$  & 2565 \\
                             & 1O Cepheids     & $-3.570\pm0.020$ & $15.969\pm0.005$  & 1682 \\ 
                             & RRab            & $-3.321\pm0.063$ & $17.440\pm0.014$  & 4378 \\ 
                             & RRc             & $-3.262\pm0.139$ & $16.986\pm0.065$  & 639  \\
                             & BL~Her          & $-2.753\pm0.403$ & $17.630\pm0.104$  & 20   \\
                             & W~Vir           & $-2.688\pm0.156$ & $17.976\pm0.164$  & 15   \\ 
                             & F-mode ACEPs    & $-2.887\pm0.140$ & $16.950\pm0.021$  & 72   \\ 
\dorule                      & 1O ACEPs        & $-3.686\pm0.275$ & $16.545\pm0.049$  & 39   \\
\hline
}

In Table~1, we present final number of objects used in this analysis
($N_{\rm fin}$) after all selection cuts and after $\sigma$-clipping
procedure (described in detail in Section~3).

\Section{Determination of the Distances to the Pulsating Stars}
All of the pulsating stars analyzed in this paper, with the exception
of peculiar W~Vir stars, follow PL relations, which allow us to
calculate their distances. However, PL relation for RV~Tau stars are
uncertain due to the large internal scatter along relation, and small
number of objects in the Magellanic Clouds. This makes the measured
distances to these stars unreliable. For this reason we decide not to
carry out the three-dimensional analysis of this T2CEPs subgroup.

We fit PL relations, separately for every group of pulsating stars
listed in Table~1. We use the reddening-independent Wesenheit index
(Madore 1976) defined as follows:
$$W_I=I-1.55\cdot(V-I).\eqno(1)$$
The 1.55 coefficient is calculated from the standard interstellar
extinction curve dependence of the {\it I}-band extinction on $E(V-I)$
reddening (Schlegel \etal 1998). Jacyszyn-Dobrzeniecka \etal (2016)
examined Wesenheit index with coefficient $1.44$ (Udalski 2003). They
found that for these two coefficients there is no significant
difference in geometry of the Magellanic System.

In this study, we marginalize the impact of metallicity ([Fe/H]) on
the PL relations. For DCEPs this is a commonly used approach as many
studies have shown that the impact of metallicity on the PL relations
is negligible (Caputo \etal 2000, Romaniello \etal 2008, Bono \etal
2008, Freedman and Madore 2011, Wielgórski \etal 2017, Gieren
\etal 2018). For RR~Lyr stars, the metallicity influences the
morphology of the horizontal branch, and so the optical PL relations
(\ie Catelan \etal 2004, Braga \etal 2015). The dependence of light
curve parameters on metallicity is significant and has been long
studied (see Skowron \etal 2016 and references therein). We compared
mean distances to the LMC and SMC calculated using both approaches --
taking into account (Jacyszyn-Dobrzeniecka 2017) and ignoring the
impact of the metallicity (from this work) on the PL relations. We
found that the difference in the distance to the LMC is at the level
of about 1\%, while for the SMC it is at level of about
3\%. Therefore, we think that distance determination without taking
into account the metallicity is sufficient for this work. It is
noteworthy, however, that while this approach is acceptable for
statistical analysis of large sample of classical pulsators, it may
cause incorrect distances to individual stars, mostly to extremely
metal-rich or extremely metal-poor objects.

\begin{figure}[b]
\vglue-4mm
\centerline{\includegraphics[width=10.6cm]{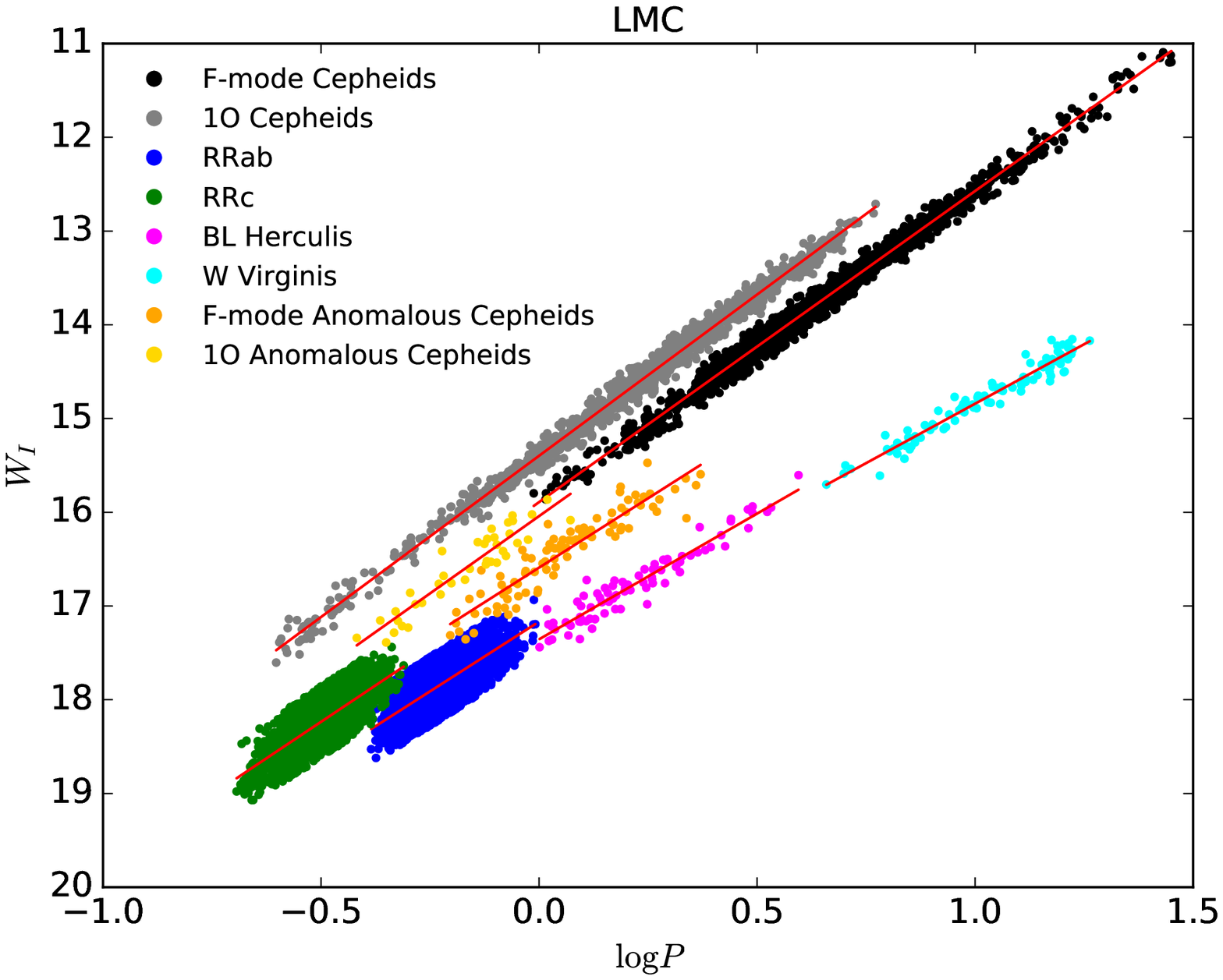}}
\centerline{\includegraphics[width=10.6cm]{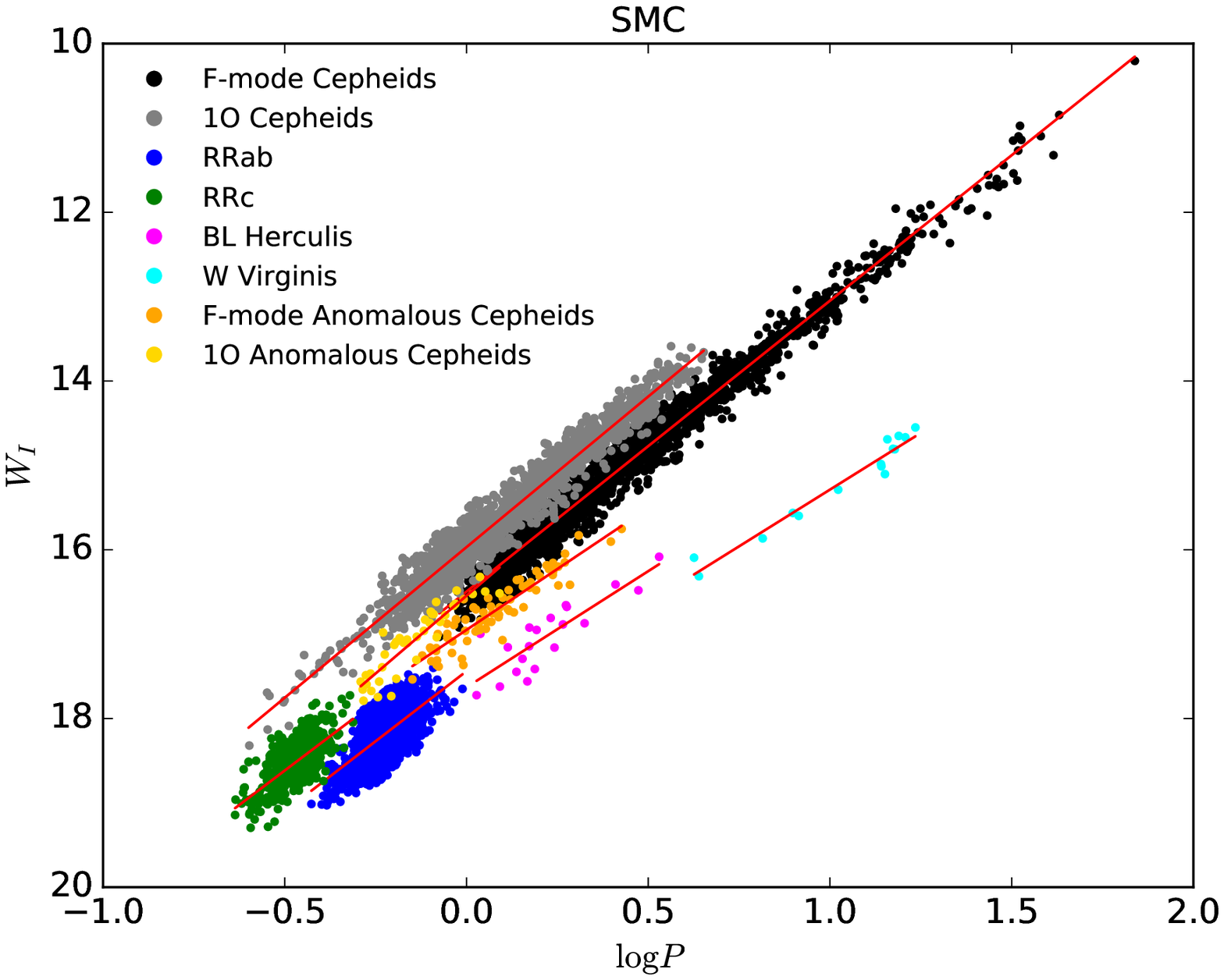}}
\FigCap{PL diagrams for DCEPs, ACEPs T2CEPs, and RR~Lyr variables in
  the Magellanic Clouds. We present PL relations after all selection
  cuts and $\sigma$-clipping procedure.}
\end{figure}
Stars that do not have measurements in both {\it I-} and {\it V-}band
are rejected from further analysis. Using the ordinary least square
method and $\sigma$-clipping procedure for each dataset we iteratively
fit a linear function in the form:
$$W_I=a\cdot\log(P)+b.\eqno(2)$$
In each iteration, we reject points deviating more than $3\sigma$ from
the predicted $W_I$ until none were rejected. The majority of the
rejected outliers are due to blending and crowding effects. The PL
diagrams for all analyzed groups of pulsating stars in the Magellanic
Clouds are shown in Fig.~3. In Table~1, we present the fitted PL
relations ($a$ and $b$ coefficients with appropriate uncertainty) for
each dataset with the remaining number of objects ($N_{\rm fin}$)
after all iterations.

We are aware that $\sigma$-clipping method is not the most appropriate for
studies of three-dimensional distributions as it was shown by Deb \etal
(2018) and Nikolaev \etal (2004), that the error distribution is not normal
for Wesenheit index at a given period (\eg due to geometry of the
Clouds). The application of this method would cause some objects that are
genuinely located closer or farther than the entire LMC/SMC sample to be
rejected as outliers. However, other studies (\ie Jacyszyn-Dobrzeniecka \etal
2016, 2017, Inno \etal 2016) proven that this method is robust enough for
studying three-dimensional samples. Therefore, we decided to use it again in
this study.

Comparing PL relations for DCEPs and RR~Lyr stars derived in this work with
that found by Jacyszyn-Dobrzeniecka \etal (2016, 2017), we can see that $a$
and $b$ coefficients marginally differ. This is mostly due to using slightly
different samples in these studies and this paper. Here we use the latest
published version of the OGLE collection of pulsating stars (Soszyñski
\etal 2017), which was updated with a number of new objects as compared to
the samples used by Jacyszyn-Dobrzeniecka \etal (2016, 2017), Additionally,
the differences in PL relations for DCEPs may be caused by the fact that
Jacyszyn-Dobrzeniecka \etal (2016) used multimode as well as single-mode
pulsators to determine PL relations, while in this study we only use
single-mode stars.

The most up-to-date PL relations for T2CEPs and ACEPs were published by
Groenewegen and Jurkovic (2017b, see Table~1). These relations are slightly
different than ours (see Table~1). The reason for this discrepancy is that we
use OGLE-IV collection of T2CEPs and ACEPs, while Groenewegen and Jurkovic
(2017b) used data from the OGLE-III phase which covered significantly smaller
area in the Magellanic System, \ie did not include the LMC northern spiral
arm. Therefore, this is the first time we present PL relations based on the
latest data.

In order to determine distances, we use the same method as
Jacyszyn-Dobrze\-niecka \etal (2016). However, we calculate distances
separately for the LMC and SMC samples. Using appropriate PL relation and
period $P$, we calculate reference Wesenheit magnitude $W_{\rm ref}$ ($a$~and
$b$~coefficients from Table~1):
$$W_{\rm ref}=a\cdot\log(P)+b.\eqno(3)$$
In the next step we calculate the relative distance modulus:
$$\delta\mu=W_I-W_{\rm ref},\eqno(4)$$
and the absolute distance given as:
$$d=d_{\rm LMC/SMC}\cdot10^{\frac{\delta\mu}{5}},\eqno(5)$$
where $d_{\rm LMC/SMC}$ are the mean distances to the LMC
($49.97\pm1.11$~kpc, Pietrzyñski \etal 2013) and SMC
($62.1\pm1.9$~kpc, Graczyk \etal 2014), respectively.

To perform statistical tests, we transform equatorial coordinates and
distances $(\alpha, \delta, d)$ to the three-dimensional Cartesian space $(x,
y,z)$ with equations given by van der Marel and Cioni
(2001) and Weinberg and Nikolaev (2001):
\setcounter{equation}{5}
\begin{equation}
\begin{aligned}
x &=-d \cdot\cos(\delta)\sin(\alpha-\alpha_{\rm cen}),\\
y &=d  \cdot(\sin(\delta)\cos(\delta_{\rm cen})-\cos(\delta)\sin(\delta_{\rm cen})\cos(\alpha-\alpha_{\rm cen})), \\
z &=d  \cdot(\cos(\delta)\cos(\delta_{\rm cen}) \cos(\alpha-\alpha_{\rm cen})+\sin(\delta)\sin(\delta_{\rm cen})).
\end{aligned}
\end{equation}
This transformation assumes that the observer is in $(0,0,0)$ and the $z$
axis is pointing toward the center of a Cloud at ($\alpha_{\rm cen}$,
$\delta_{\rm cen}$). We transform LMC and SMC stars coordinates separately
using the centers of the Magellanic Clouds based on the RR~Lyr variables
distributions (Jacyszyn-Dobrzeniecka \etal 2017):

\begin{equation}
\begin{aligned}
(\alpha_{\rm cen, LMC},\delta_{\rm cen, LMC})=(5\uph21\upm31\zdot\ups2, -69\arcd36\arcm36\arcs),\\
(\alpha_{\rm cen, SMC},\delta_{\rm cen, SMC})=(0\uph55\upm48\zdot\ups0, -72\arcd46\arcm48\arcs).
\end{aligned}
\end{equation}

To visualize and compare by eye stars distributions we use two-dimensional
sky map in Hammer equal-area projection. We transform equatorial coordinates
$(\alpha, \delta)$ to coordinates in the Hammer system ($x_{\rm Hammer},
y_{\rm Hammer})$. In this coordinates transformation, the $z$ axis is
pointing toward $(\alpha_{\rm cen}, \delta_{\rm cen})=(3\uph20\upm,
-72\arcd)$. Hammer coordinates are calculated as follow:
\begin{equation}
\begin{aligned}
\alpha_{\rm b} &= \alpha+\left(\frac{\pi}{2}-\alpha_{\rm cen}\right),\\
l              &= \arctan\left(\frac{\sin(\alpha_{\rm b})\cos(\delta_{\rm cen)}+\tan(\delta)\sin(\delta_{\rm cen})}{\cos(\alpha_{\rm b})}\right)-\frac{1}{2}\pi,\\
\beta          &= \arcsin\left(\sin(\delta) \cos(\delta_{\rm cen})-\cos(\delta) \sin(\delta_{\rm cen}) \sin(\alpha_{\rm b})\right), \\
x_{\rm Hammer} &=-\frac{2\sqrt2\cos(\beta) \sin(\frac{l}{2})}{\sqrt{1+\cos(\beta) \cos(\frac{l}{2})}},\\
y_{\rm Hammer} &= \frac{ \sqrt2\sin(\beta)}{\sqrt{1+\cos(\beta)\cos(\frac{l}{2})}}.
\end{aligned}
\end{equation}

Eqs.(6) and (8) are based on Eqs.(7--14) from Jacyszyn-Dobrzeniecka (2016)
though we applied a small correction to one of them (there was a typo in
Eq.(8) and a coefficient of $-\frac{1}{2}\pi$ was missing in the right hand
side). The correct version that we use here is presented above.

\Section{Method of Performing Statistical Tests}
In order to compare three-dimensional distributions of pulsating stars from
our collection we use the extended Kolmogorov-Smirnov (KS) statistical
test. The two-sample two-dimensional version of the KS test was introduced by
Peacock (1983), who found that this test is almost independent from the
distribution type. This means that for large samples, critical values of the
test statistics should not differ significantly. Gosset (1987) has extended
the Peacock's idea to three-dimensional distributions. The test statistic
$D_n$ is defined as the maximum absolute difference between two cumulative
distribution functions. In three-dimensional space, 8 pairs of cumulative
distribution functions are needed to calculate the test statistic
$D_n$. Peacock (1983) pointed out that it was better to work with the test
statistic $Z_n$ defined as follow:
$$Z_n=\sqrt{\frac{n_1 n_2}{n_1+n_2}}D_n\eqno(9)$$
where $n_1$ and $n_2$ are the sample sizes. Assuming that the first sample
comes from the $F(x,y,z)$ distribution, and the second sample comes from the
$G(x,y,z)$ distribution, we test the null ($H_0$) and alternative ($H_1$)
hypotheses defined as follow:
$$H_0: F(x,y,z) = G(x,y,z), \qquad H_1: F(x,y,z) \neq G(x,y,z)\eqno(10)$$
where $H_0$ means that the analyzed samples come from the same distribution,
while $H_1$ means that samples come from different distributions.

The Peacock's idea of the multidimensional, two-sample Kolmogorov-Smirnov
test is implemented in the statistical software R (R Core Team -- {\it
  https://www.R-project.org}). The package named {\sf Peacock.test} (Xiao
2017 and {\it https://CRAN.R-project.org/package=Peacock.test}) allows
calculating the test statistic $D_n$ for two samples in two- and
three-dimensional spaces. The test statistic $D_n$ obtained in this
calculation is converted to $Z_n$ using the sample sizes $n_1$ and
$n_2$. Peacock (1983) determined that sizes of the tested samples $n_1$~and
$n_2$ should be greater than 10, while Gosset (1987) indicated that the
sample sizes must be greater than~5.

The main goal of this study is to examine similarities between spatial
distributions of classical pulsators (DCEPs and RR~Lyr variables) and T2CEPs
or ACEPs. These similarities would allow us to better understand the nature
of these stars, especially their possible evolutionary histories. We test all
possible combinations of pairs of samples. Typically, the first sample with
size $n_1$ contains T2CEPs or ACEPs, while the second sample with size $n_2$
consists of other classical pulsators. The groups of DCEPs or RR~Lyr stars
are much larger than other groups of pulsating stars. Therefore, we draw
without returning a set of stars from classical pulsator distributions with
size $n_2$, which is three times larger than $n_1$. We draw 1000 such
samples, for which we compute the test statistics $D_n$, and later we convert
it to $Z_n$. The statistical test for large samples requires a large amount
of computational time, so we decided to count the tests repeatedly for
smaller samples.

In their papers, Peacock (1983) and Gosset (1987) give empirical formulae for
estimating probability $p$ and testing the hypotheses. In the main part of
our analysis, we use two-sided critical values of the ``theoretical'' test
statistic $Z_n$ for testing hypotheses and making decision whether to reject
or accept $H_0$. However, we compare our decision-making method with
asymptotic equations given by Gosset (1987). We discuss a comparison of these
two methods in Section~6.

To test our hypotheses, we have to use the ``theoretical'' distribution of
the test statistic in order to be able to compare this distribution (and its
critical values) to the distribution of the test statistic of the tested
samples. Bearing in mind the distribution-independent property of this test,
we decide to use different ``theoretical'' distribution for each tested pair
of samples. The sizes of the tested samples are not so large, therefore, the
critical values of the test statistics are slightly different in each
cases. Taking advantage of the fact that our collection of pulsating stars is
very large, we decided to build a ``theoretical'' distributions of the test
statistics based on all DCEPs and RR~Lyr stars. For example, to test the
hypotheses for BL~Her ($n_1=79$) and RR~Lyr variables ($n_2=3n_1=237$), we
need ``theoretical'' distribution based only on RR~Lyr stars. Hence, we use
RR~Lyr stars distribution to draw without returning samples with sizes $n_1$
and $n_2$, which were tested 5000 times. We assume to test the hypotheses at
the significance level of $\alpha = 0.05$. Therefore, for each
``theoretical'' distribution of $Z_n$ we calculate 2.5th and 97.5th
percentiles. The regions with $Z_n$ smaller or equal than 2.5th percentile
and larger or equal than 97.5th percentile are the critical region of the
test statistic (region of the $H_0$ rejection).

In the next step, we compare each $Z_n$ calculated for our pulsating stars
with critical values of the ``theoretical'' test statistic distribution. If
the value of $Z_n$ is in the critical region of the test statistic we
conclude, that there are clear grounds to reject the null hypothesis $H_0$,
so we have to accept alternative hypothesis $H_1$. However, when value $Z_n$
is between $2.5$th and $97.5$th percentiles we can conclude, that there is no
sufficient evidence to reject the null hypothesis $H_0$, so we accept $H_0$
(hence, we called this region acceptability region). We assumed that there is
no evidence to reject $H_0$ (and to indicate the possible similarity between
two distributions) if at least 90\% of the $Z_n$ are in the region between
critical values of the test statistics. Hence, we think that if over 900 test
statistics are outside the rejection area, it is likely that we will be able
to find similarity in the spatial distributions of stars.

In the current framework of stellar pulsations and evolution theories, the
spatial distributions of classical pulsators should not depend on the
pulsation mode, and so in our statistical tests we did not divide DCEPs,
RR~Lyr stars and ACEPs into smaller groups with distinct pulsation
modes. Looking at the sky distributions of classical pulsating stars, the
differences between F-mode and 1O distributions are noticeable. However, our
investigation shows that these differences are relatively small and they do
not affect results of statistical tests. For instance, while testing BL~Her
stars with RR~Lyr variables, the percentage of the test statistics in the
acceptability region is 91.80\% when RRab and RRc stars are treated as one
group, but in case of testing BL~Her stars with RRab and RRc stars
separately, there is 92.50\% and 93.00\% of $Z_n$ in the acceptability
region, respectively. We obtained very similar results for the tests W~Vir
stars with DCEPs. The percentage of the test statistics outside the critical
region of the test is 97.10\% when F-mode and 1O DCEPs are treated as a
single group, and 93.90\% and 90.70\% when they are treated separately. These
differences are very similar for the SMC stars. In the case of remaining
pairs, the null hypothesis is rejected because it does not meet our
acceptability criterion.

\vspace*{-9pt}
\Section{Results of our Analysis}
\vspace*{-5pt} In Tables~2 and 3, we present the results of our statistical
tests for all tested pairs of samples in the LMC and SMC. Pairs of
pulsating classes for which we can conclude similarity in their spatial
distributions are marked in bold. All distributions of the test statistic
$Z_n$ in comparison to the ``theoretical'' distributions of the test
statistic of the RR~Lyr stars and DCEPs are presented in Fig.~4 for the LMC
(two left panels) and for the SMC (two right panels). Using solid lines, we
marked from the left 2.5th, 50th (median), and 97.5th percentiles. The
areas marked in gray are the rejection areas of the null hypothesis $H_0$.

In Figs.~5--7, we present spatial distributions of T2CEPs and ACEPs in
comparison to the DCEPs and RR~Lyr variables distributions. We present the
most interesting cases only for which results are similar in both Magellanic
Clouds. In Figs.~5--7, the top plot presents two-dimensional distributions of
pulsating stars in the Magellanic Clouds in the equal-area Hammer projection
(the coordinates transformation described in Section~3). The middle and the
bottom plots present $xy$ (face-on view), $xz$ (plan view) and $yz$ (edge-on
view) planes for the LMC and the SMC, respectively. To estimate shapes of the
galaxies in each projection we use a standard kernel density estimation (KDE)
based on the DCEPs or RR~Lyr stars. The densest areas are marked in navy
blue. Additionally, in each projection we plot the normalized density
contours.

The numbers of T2CEPs in the SMC are significantly smaller than in the LMC,
therefore fitted PL relations have larger uncertainties (see
Table~1). Taking it into consideration, our tests may not be very accurate
(despite the lower limit on the samples sizes, Peacock 1983, Gosset 1987,
which is 10, and 5, respectively). Due to these reasons, we think that our
analysis should be based mainly on the LMC, while SMC should be treated as an
additional clue, not the foundation to drawing conclusions.

\subsection{BL~Her}
Statistical tests in the LMC indicate that the spatial distribution of BL~Her
and RR~Lyr stars are similar (Fig.~4b, LMC). There is $91.80\%$ of the test
statistics in the acceptability region. This result is not surprising,
because we expect that BL~Her stars are old and have ages similar to the
RR~Lyr variables. Moreover, masses of the BL~Her and RR~Lyr stars are
comparable (Groenewegen and Jurkovic 2017ab). In Fig.~5, we can see that some
BL~Her stars in the LMC are concentrated in the center of the galaxy, but the
vast majority of these objects are distributed out of the center creating a
vast halo, which is typical for old populations. In the SMC, the statistical
tests for BL~Her stars with RR~Lyr variables give a comparable result
(Fig.~4b, SMC). However, the difference is in the tests with DCEPs, for which we
conclude similarities (Fig.~4a, SMC). As we mentioned before, it seems that this
is due to a small number of T2CEPs in the SMC. Moreover, it is difficult to
compare by eye these two distributions in the SMC with such small number of
BL~Her variables.
\begin{figure}[t]
\centerline{\includegraphics[width=6.7cm, bb=10 40 410 720, clip=]{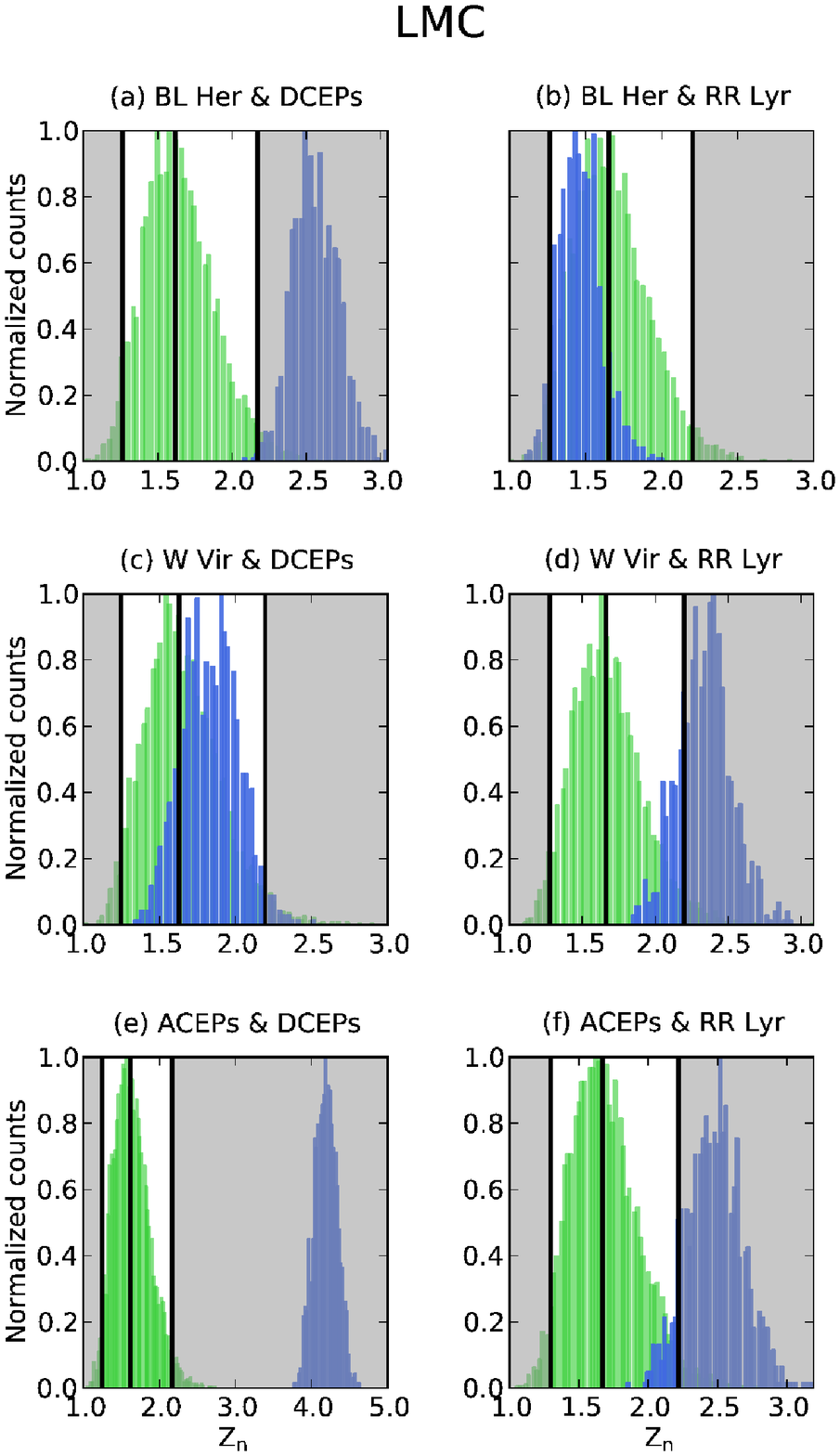}
  \includegraphics[width=6.7cm, bb=10 40 410 720, clip=]{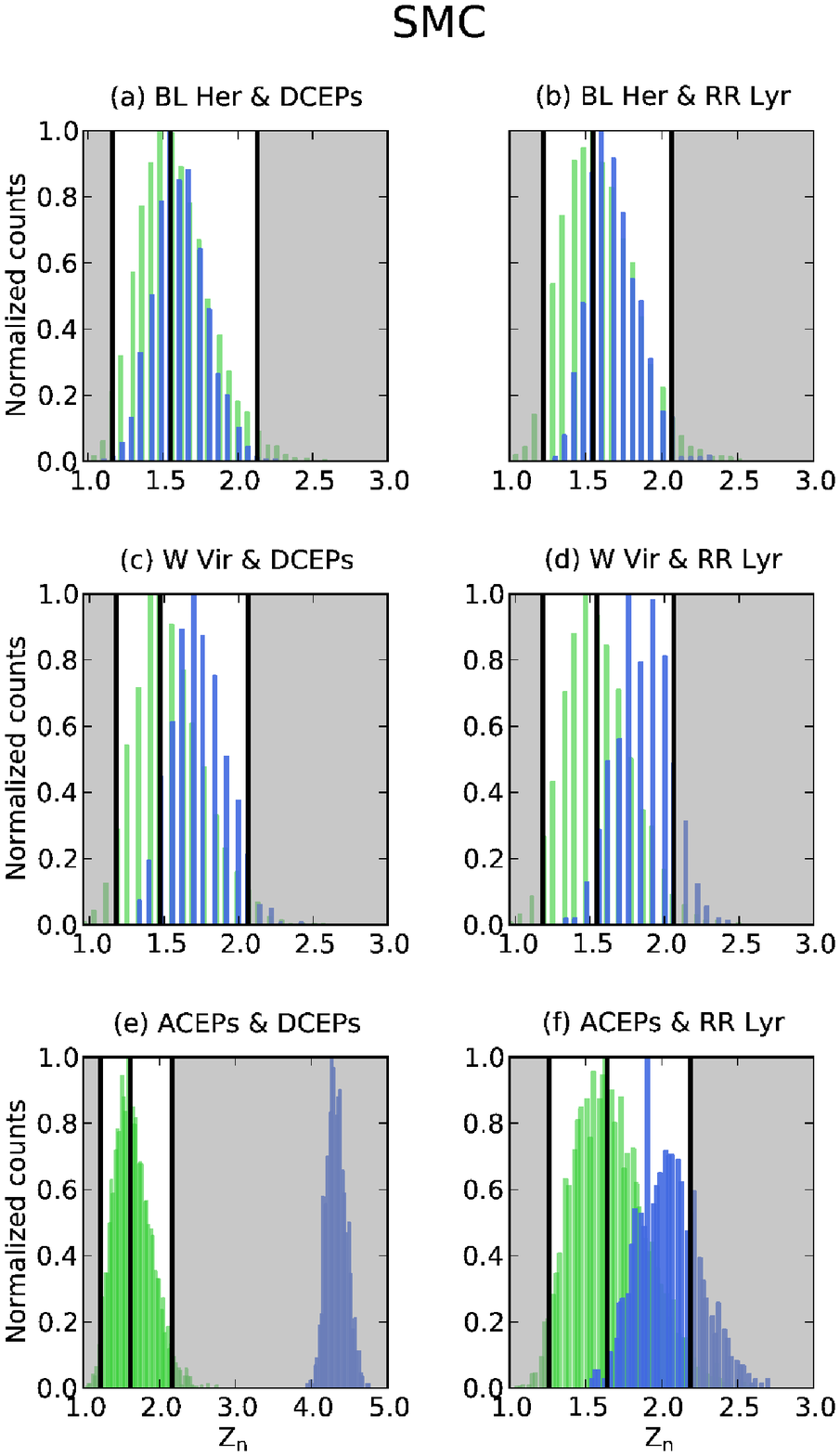}}
\vskip7pt 
\FigCap{Distributions of the test statistic $Z_n$ for each tested
  pair of samples from the LMC ({\it two left panels}) and SMC ({\it two
    right panels}). With green color we marked ``theoretical''
  distributions of the test statistic built based on DCEPs and RR~Lyr
  stars, while blue histograms show the test statistic obtained during the
  tests of spatial distributions of T2CEPs and ACEPs with other classical
  pulsators. With solid, black lines we marked percentiles 2.5th, 50th
  (median), and 97.5th (from left side, respectively). The area marked in
  gray is the null hypothesis rejection area (critical region of the test
  statistic). Each plot represents different pair of tested samples.}
\end{figure}

\begin{figure}[p]
\centerline{\includegraphics[width=13cm]{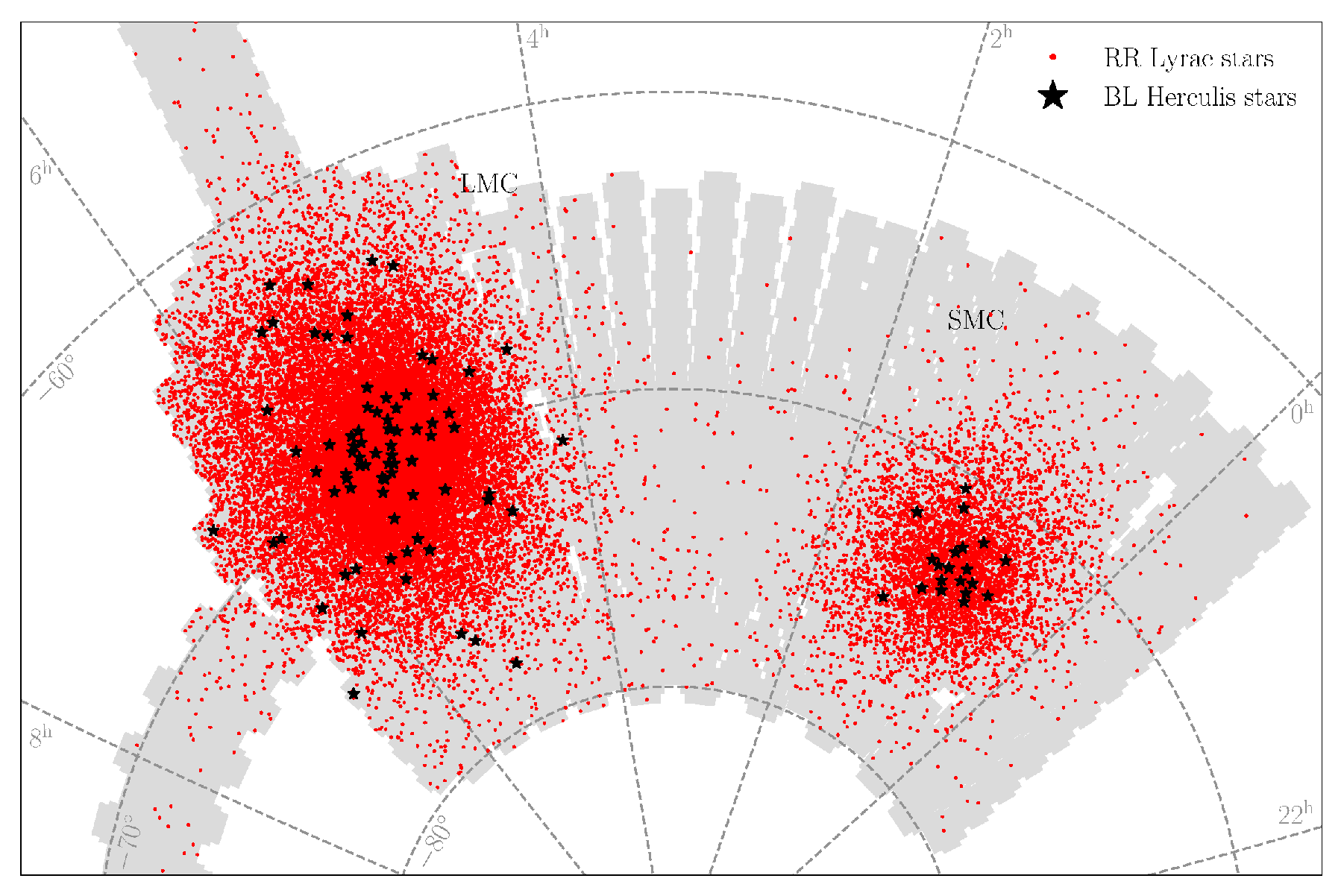}}
\vskip2mm
\centerline{\includegraphics[width=13cm]{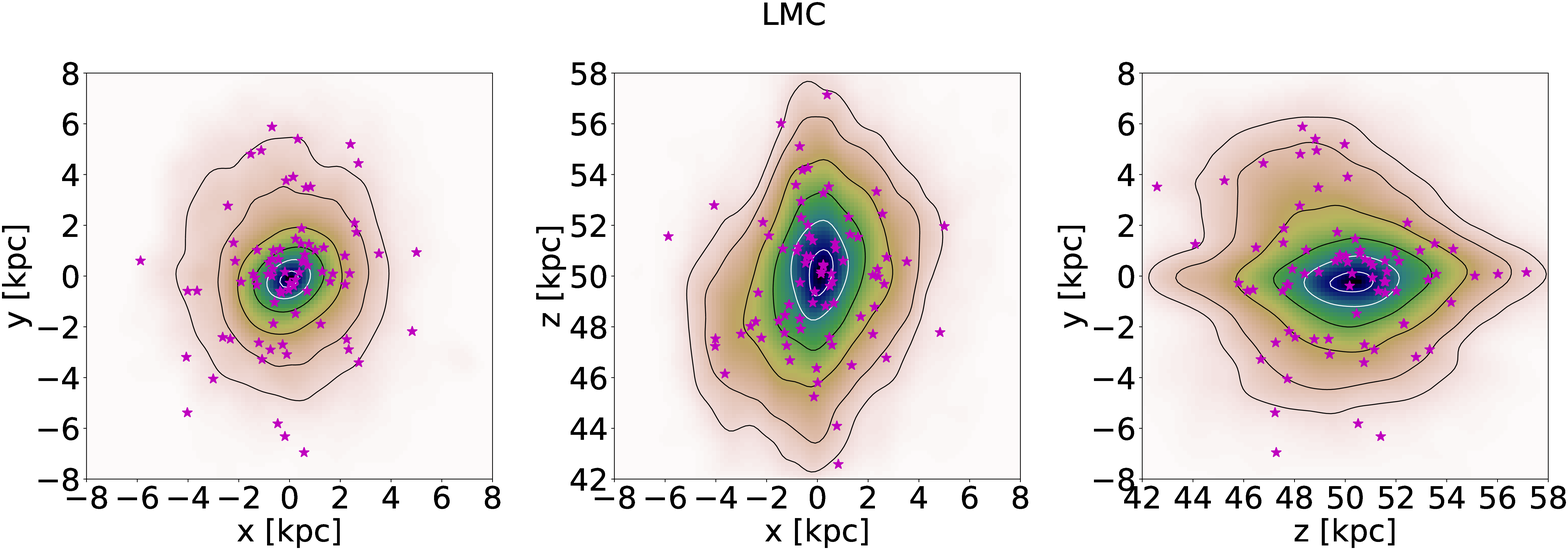}}
\vskip2mm
\centerline{\includegraphics[width=13cm]{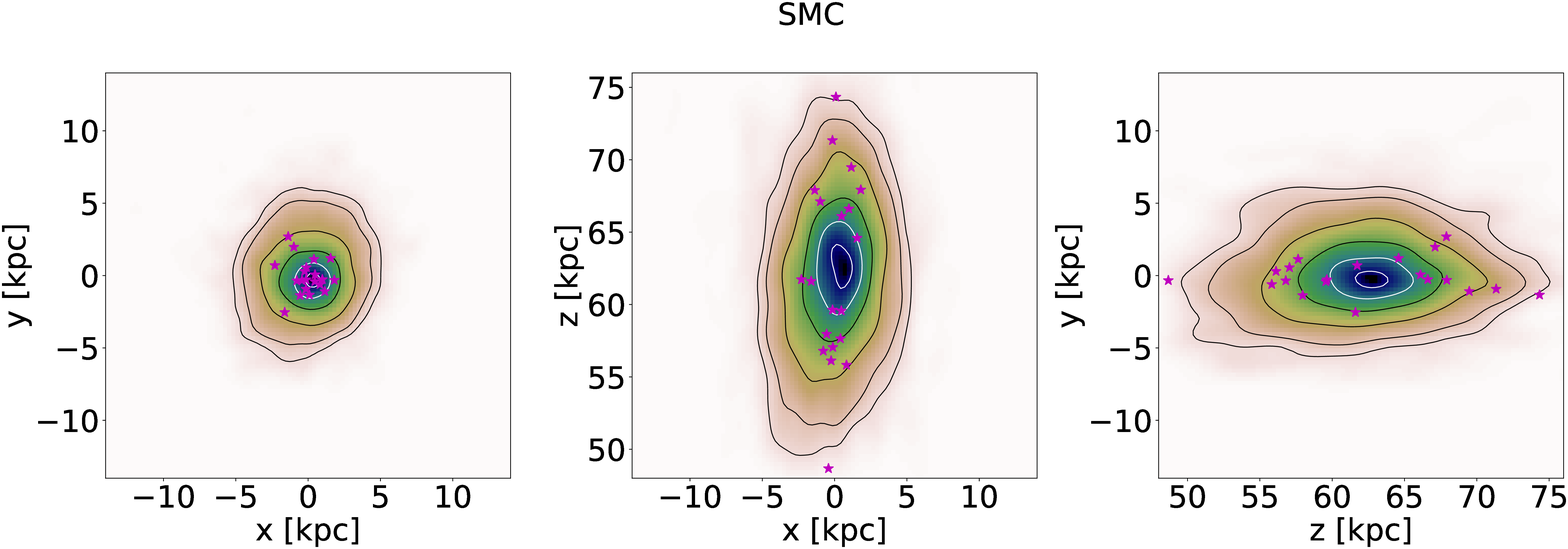}} 
\vskip2pt
\FigCap{Spatial distribution of BL~Her stars in comparison to the
distribution of RR~Lyr stars. {\it Top panel} presents equal-area Hammer
projection of the Magellanic System. Area marked in gray is the OGLE-IV
footprint. {\it Middle and bottom panels} present Cartesian projections
for the LMC and the SMC, respectively. We estimate shapes of the galaxies
in each projection using standard kernel density estimation (KDE) and
RR~Lyr stars, which are marked with color map. Additionally, in each
projection we plot normalized density contours. From the center of the
galaxies to the edge, we plot normalized density with value: 95\% (first
white contour), $75\%$ (second white contour), 50\%, 25\%, 10\% and
5\%. With magenta points, we marked BL~Her stars.}
\end{figure}
\begin{figure}[p]
\centerline{\includegraphics[width=13cm]{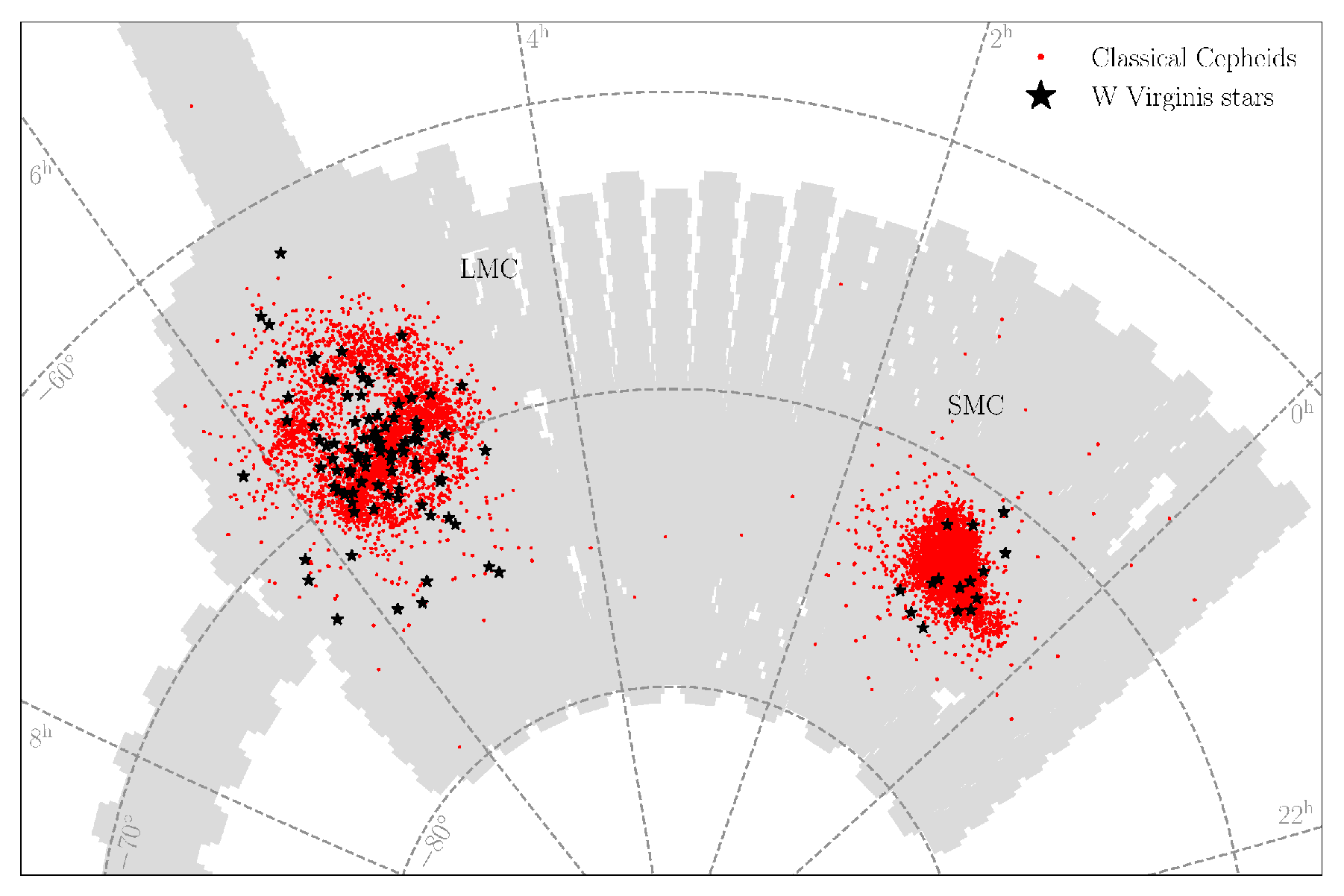}}
\vskip5mm
\centerline{\includegraphics[width=13cm]{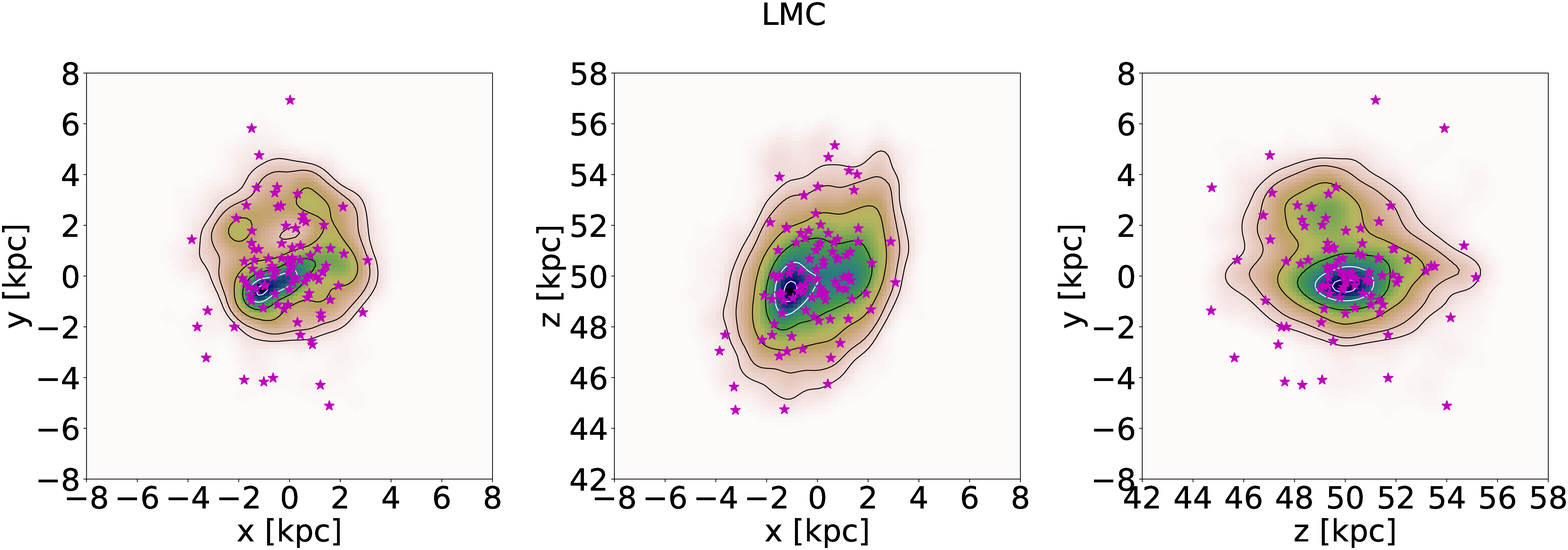}}
\vskip5mm
\centerline{\includegraphics[width=13cm]{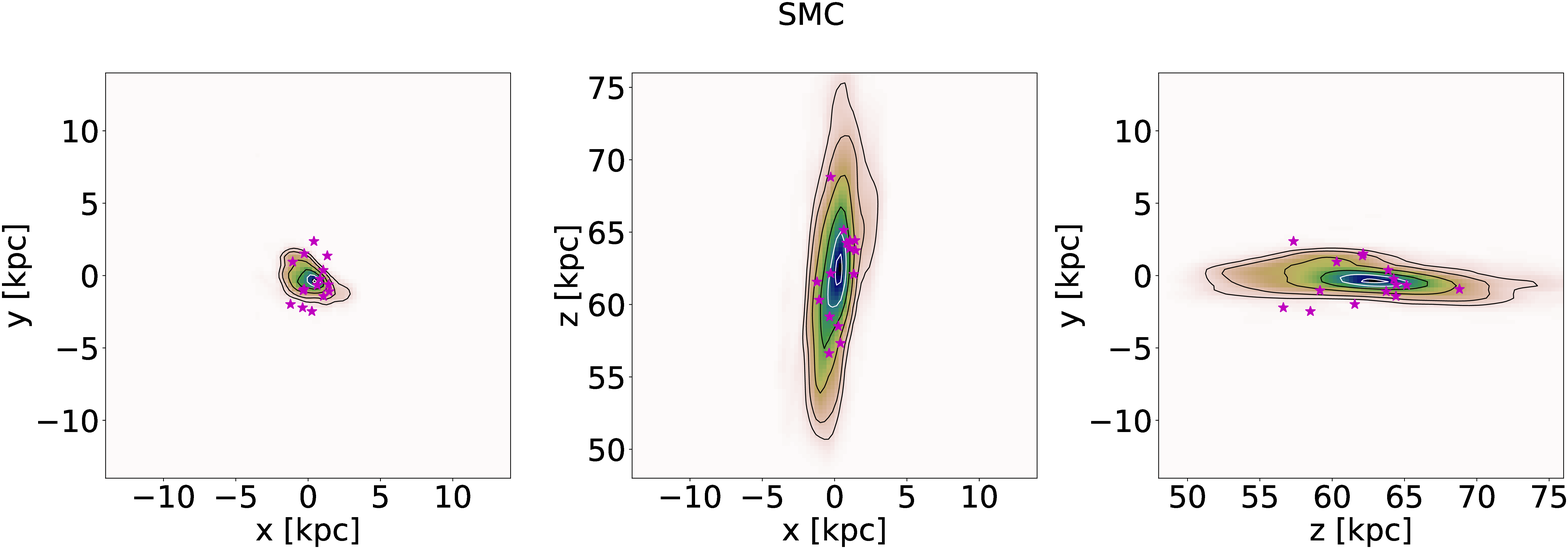}} 
\vskip5pt
\FigCap{Same as Fig.~5, but for W~Vir stars and DCEPs.}
\end{figure}

\begin{figure}[p]
\centerline{\includegraphics[width=13cm]{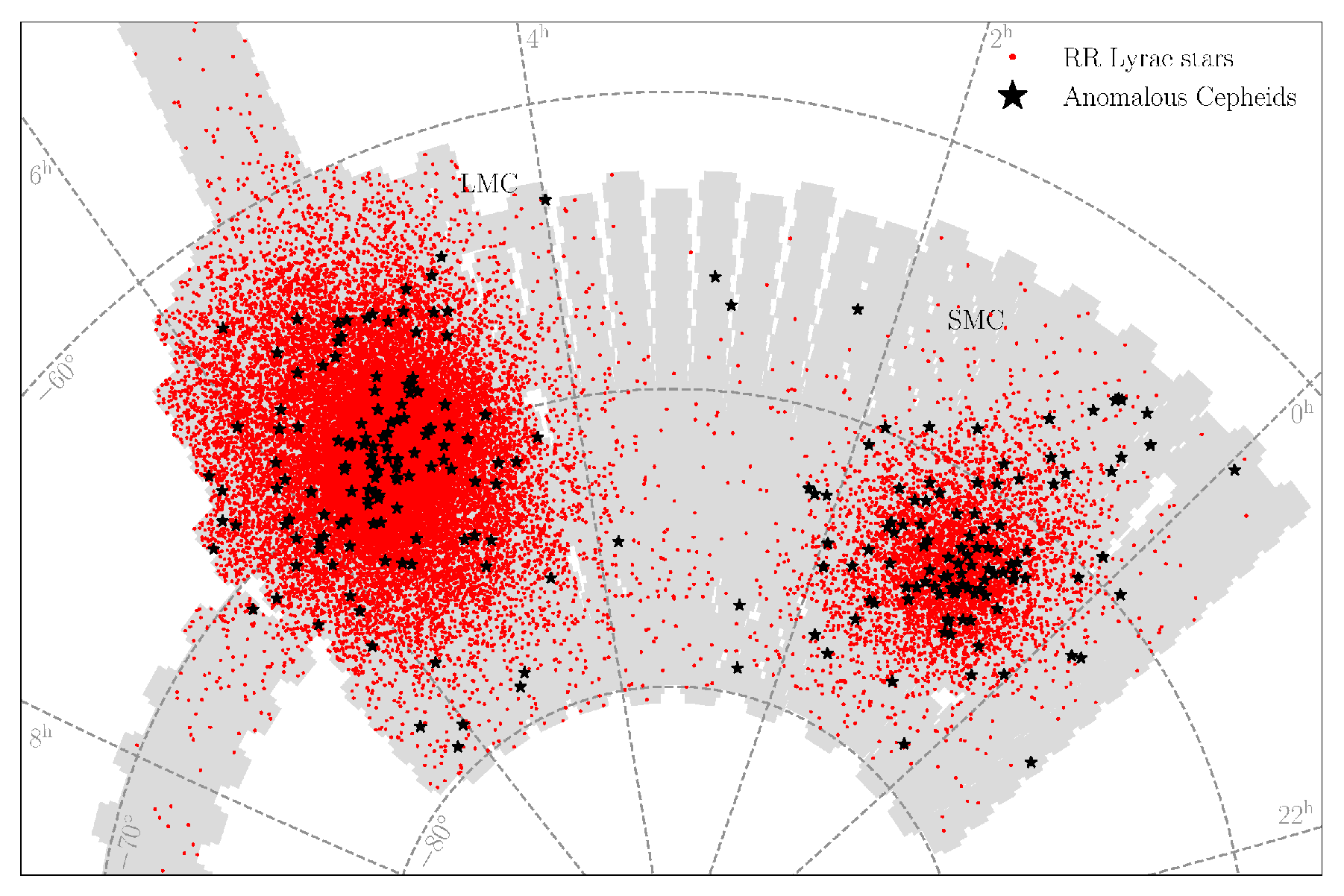}}
\vskip5mm
\centerline{\includegraphics[width=13cm]{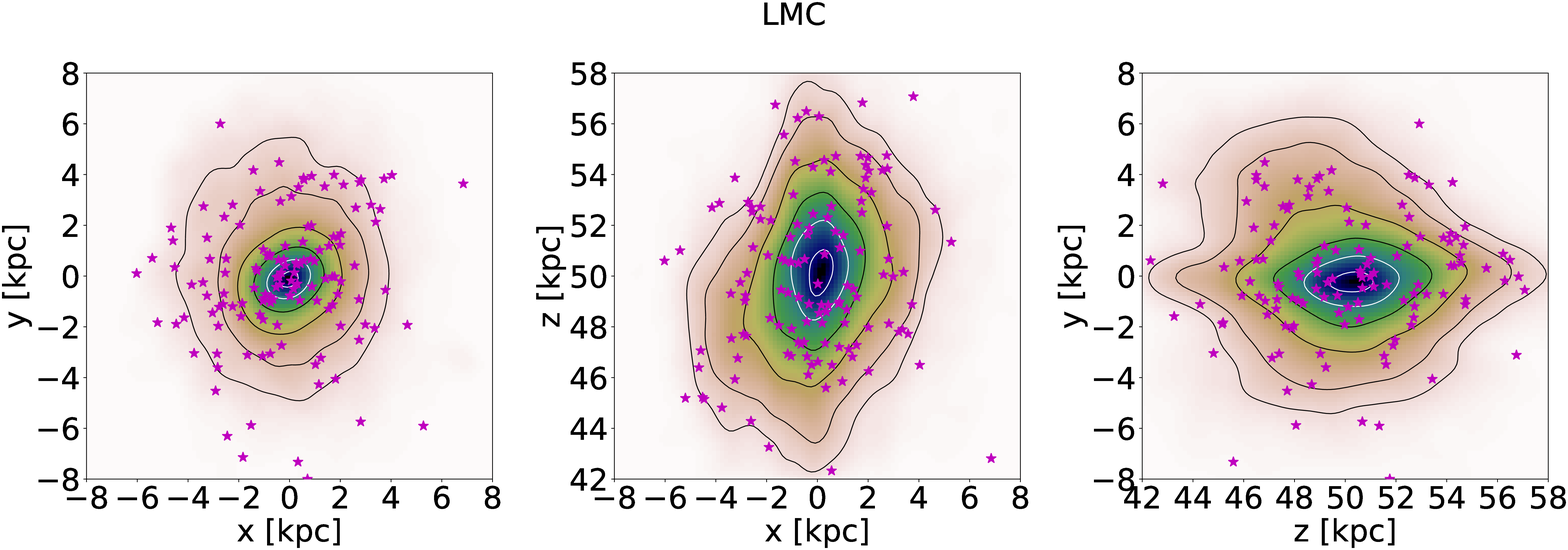}}
\vskip5mm
\centerline{\includegraphics[width=13cm]{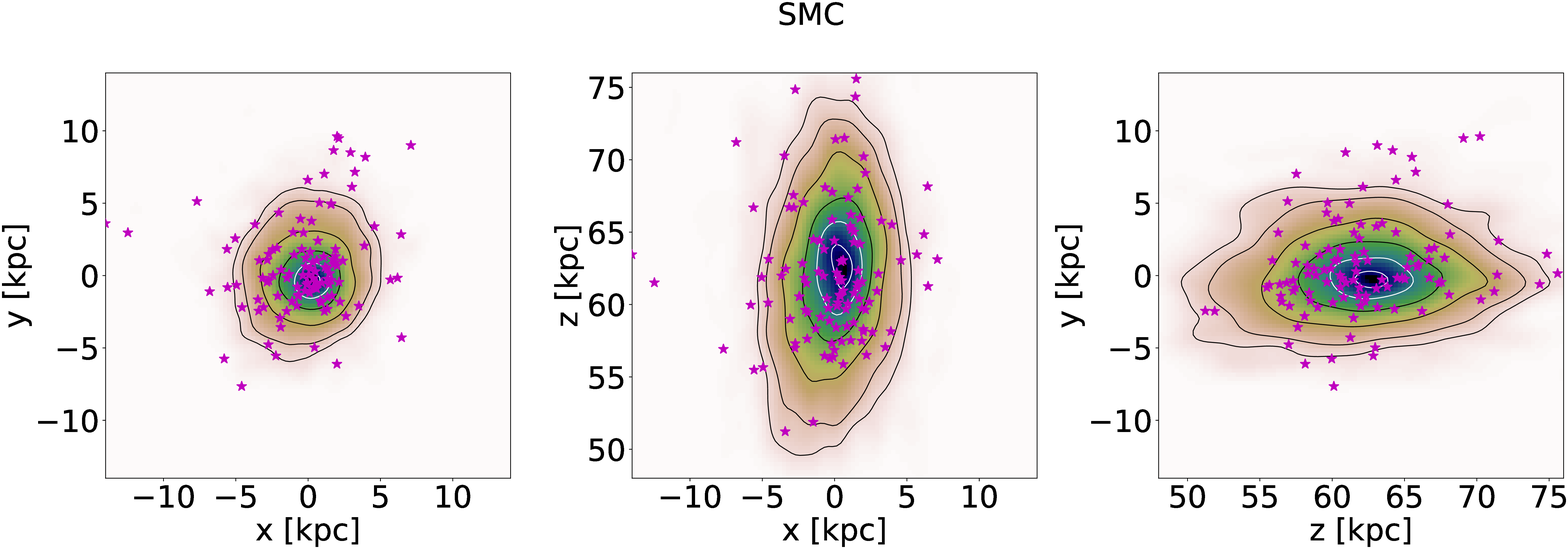}} 
\vskip5pt
\FigCap{Same as Fig.~5, but for ACEPs and RR~Lyr stars.}
\end{figure}
\newpage
\begin{landscape}
\MakeTableee{l|c|c|c|c|c|c|c|r}{19cm}{Results of the statistical tests in the Large Magellanic Cloud}
{\hline
\multirow{3}[6]{*}{tested samples ($n_1$ and $n_2$)} & \multicolumn{8}{c}{LMC} \\
\cline{2-9}                                          & \multicolumn{2}{c|}{sample sizes} & \multicolumn{3}{c|}{,,theoretical'' test statistics} & \multicolumn{2}{c|}{test statistics of the tested samples} & \multirow{2}[4]{*}{decision} \\
\cline{2-8}                                          & $n_1$                             & $n_2$                & 2.5th & 50th  & 97.5th& $\% Z_n$ in crit. region & $\% Z_n$ in accept. region\\ 
\hline
BL~Her and DCEPs (Fig.~4a, LMC)                      & \multirow{2}{*}{79}               & \multirow{2}{*}{237} & 1.266 & 1.624 & 2.176 & 99.80      & 0.20 & $H_0$ rejected \\
{\bf BL~Her and RR~Lyr stars (Fig.~4b, LMC)}         &                                   &                      & 1.267 & 1.656 & 2.209 & {\bf 8.20} & {\bf 91.80} & {\bf $H_0$ accepted} \\
\hline
{\bf W~Vir and DCEPs (Fig.~4c, LMC)}                 & \multirow{2}{*}{94}               & \multirow{2}{*}{282} & 1.251 & 1.638 & 2.203 & {\bf 2.90} & {\bf 97.10} & {\bf $H_0$ accepted} \\ 
W~Vir and RR~Lyr stars (Fig.~4d, LMC)                &                                   &                      & 1.280 & 1.667 & 2.203 & 80.30      & 19.70       & $H_0$ rejected \\ 
\hline
ACEPs and DCEPs (Fig.~4e, LMC)                       & \multirow{2}{*}{133}              & \multirow{2}{*}{399} & 1.252 & 1.627 & 2.178 & 100.00     & 0.00        & $H_0$ rejected \\ 
ACEPs and RR~Lyr stars (Fig.~4f, LMC)                &                                   &                      & 1.301 & 1.677 & 2.228 & 92.50      & 7.50        & $H_0$ rejected \\ 
\hline}

\MakeTableee{l|c|c|c|c|c|c|c|r}{19cm}{Results of the statistical tests in the Small Magellanic Cloud}
{\hline
\multirow{3}[6]{*}{tested samples ($n_1$ and $n_2$)} & \multicolumn{8}{c}{SMC} \\
\cline{2-9}                                          & \multicolumn{2}{c|}{sample sizes} & \multicolumn{3}{c|}{,,theoretical'' test statistics} & \multicolumn{2}{c|}{test statistics of the tested samples} & \multirow{2}[4]{*}{decision} \\
\cline{2-8}                                          & $n_1$                             & $n_2$                & $2.5$th & $50$th  & $97.5$th           & $\% Z_n$ in crit. region & $\% Z_n$ in accept. region  \\ \hline
{\bf BL~Her and DCEPs (Fig.~4a, SMC)}                & \multirow{2}{*}{20}               & \multirow{2}{*}{60}  & 1.162   & 1.549   & 2.130 & {\bf 0.70} & {\bf 99.30} & {\bf $H_0$ accepted}\\
{\bf BL~Her and RR~Lyr stars (Fig.~4b, SMC)}         &                                   &                      & 1.226   & 1.549   & 2.066 & {\bf 3.10} & {\bf 96.90} & {\bf $H_0$ accepted}\\ \hline
{\bf W~Vir and DCEPs (Fig.~4c, SMC)}                 & \multirow{2}{*}{15}               & \multirow{2}{*}{45}  & 1.193   & 1.491   & 2.087 & {\bf 5.50} & {\bf 94.50} & {\bf $H_0$ accepted}\\ 
W~Vir and RR~Lyr stars (Fig.~4d, SMC)                &                                   &                      & 1.193   & 1.565   & 2.087 & 16.50      & 83.50       & $H_0$ rejected \\ 
\hline
ACEPs and DCEPs (Fig.~4e, SMC)                       & \multirow{2}{*}{111}              & \multirow{2}{*}{333} & 1.233   & 1.617   & 2.165 & 100.00     & 0.00        & $H_0$ rejected \\ 
ACEPs and RR~Lyr stars (Fig.~4f, SMC)                &                                   &                      & 1.260   & 1.644   & 2.192 & 25.90      & 74.10       & $H_0$ rejected \\  
\hline
\noalign{\vskip6pt}
\multicolumn{9}{p{19cm}}{Pairs of pulsating stars for which we can conclude
  similarity in their spatial distributions in Tables~2 and 3 are marked in
  bold}
}
\end{landscape}

\subsection{W~Vir}
\begin{figure}[b]
  \centerline{\includegraphics[width=12.5cm]{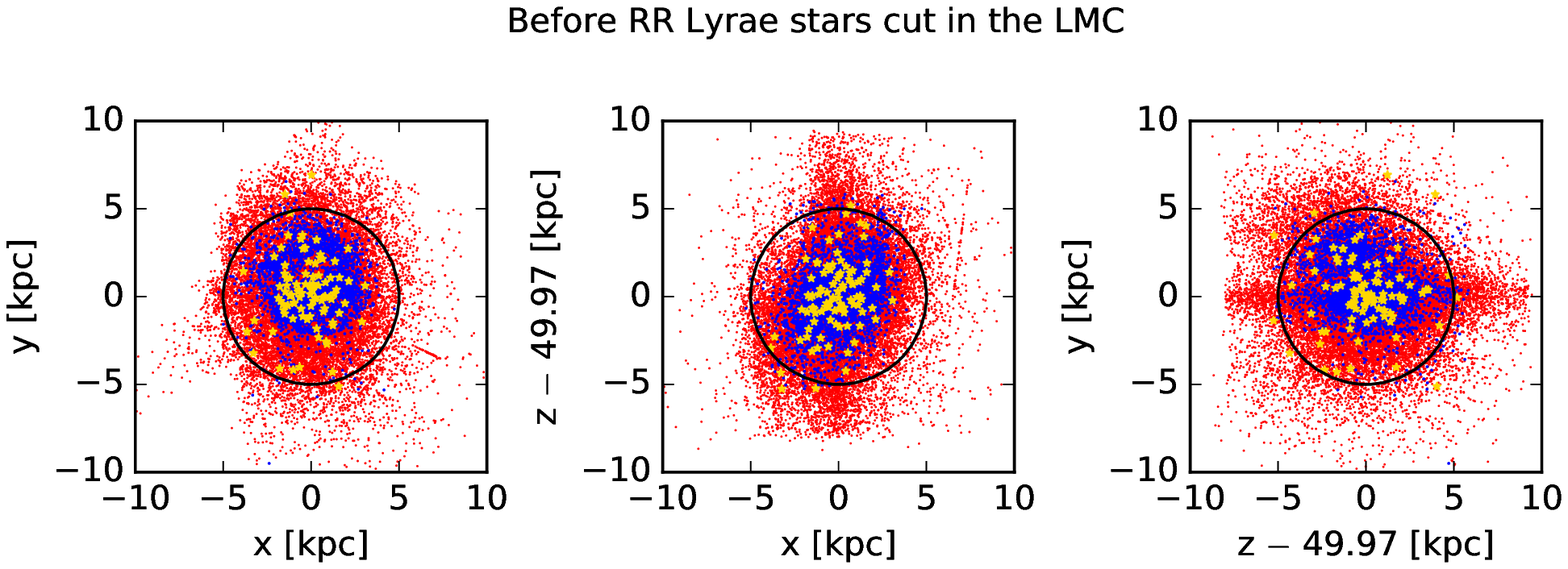}}
  \centerline{\includegraphics[width=12.5cm]{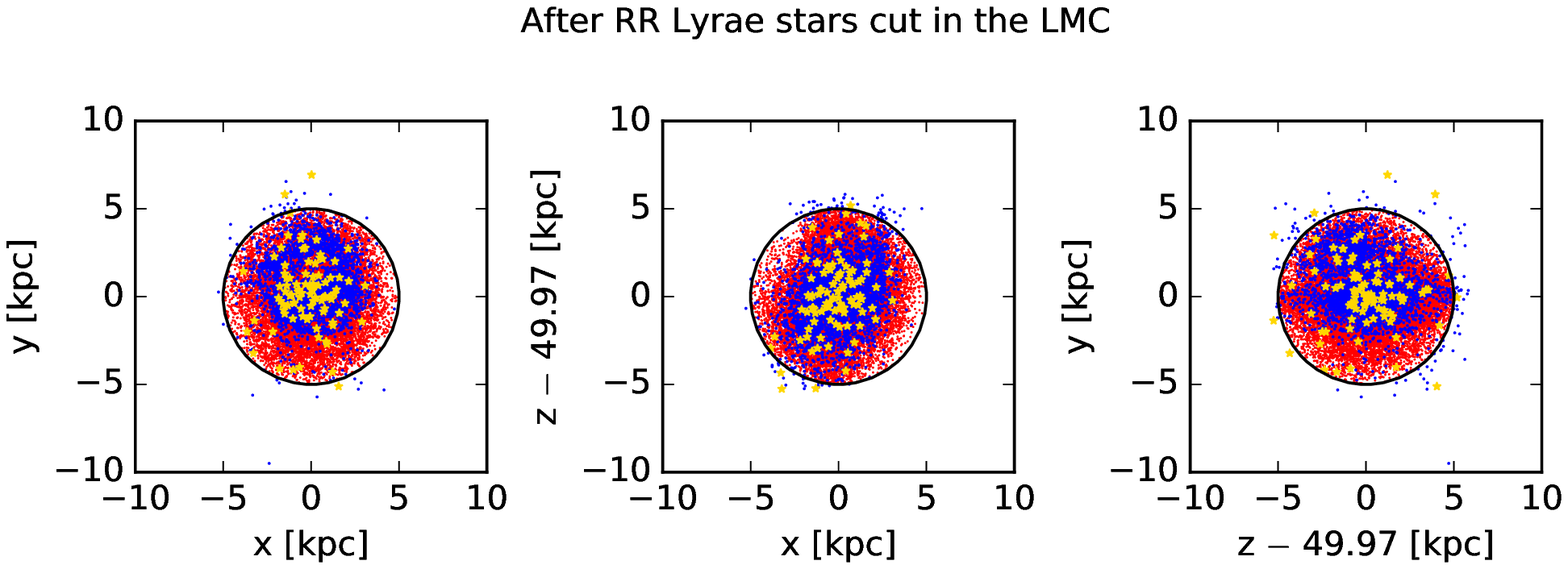}} 
\FigCap{Cutting procedure for RR~Lyr outskirts regions in the Large
  Magellanic Cloud. With red dots we marked RR~Lyr stars, blue dots
  correspond to DCEPs, while gold stars present W~Vir
  distribution. Black circles represent the cut boundary of the RR~Lyr
  stars halo, at the radius of $R=5$~kpc.}
\end{figure}
Our statistical tests show that spatial distribution of W~Vir stars is
comparable to DCEPs in both Magellanic Clouds. The vast majority of the
test statistics outside the rejection areas indicates that W~Vir variables
follow the distribution similar to that of young stars (Fig.~4c, LMC and
SMC). The similarity of these two distributions is difficult to see in
Fig.~6, taking into account the fact that some of these stars are located
in the area where there is a halo which consists of the old stars, and
where young stars are practically absent. In the tests of W~Vir stars with
RR~Lyr variables in both Clouds (Fig.~4d, LMC and SMC), some of the test
statistics are outside of the rejection area. In the LMC it is 19.70\%,
whereas in the SMC it is 83.50\% . Therefore, the most important question
is whether the similarity between W~Vir stars and DCEPs is related to the
structures created by young population stars, \ie the bar and spiral
arms. The simplest test that can verify this thesis is to compare the
spatial distribution of W~Vir stars with RR~Lyr variables, but with the
limitation of the area in the three-dimensional space occupied by RR~Lyr
stars to the area occupied by DCEPs. We limit RR~Lyr variables outskirts in
the LMC to a sphere with a radius $R=5$~kpc, where the vast majority of
DCEPs are located. In Fig.~8 we present RR~Lyr stars before and after
limitation. Then, we compared W~Vir stars distribution with the
distribution of RR~Lyr variables. In Fig.~9, we present ``theoretical''
distribution in comparison to the test statistic distribution for W~Vir and
RR~Lyr stars. After RR~Lyr halo cut, 94.80\% of test statistics are outside
the rejection area. This result suggests that W~Vir distribution
similarities to DCEPs is not directly related to the structures created by
the young population, but it is related to the lack of a big halo that
RR~Lyr stars form. Therefore, we are able to conclude that W~Vir variables
are intermediate-age stars with ages somewhere between DCEPs and RR~Lyr
variables, or they could be a mixture of an old and intermediate-age stars.
\begin{figure}[htb]
\centerline{\includegraphics[width=5cm]{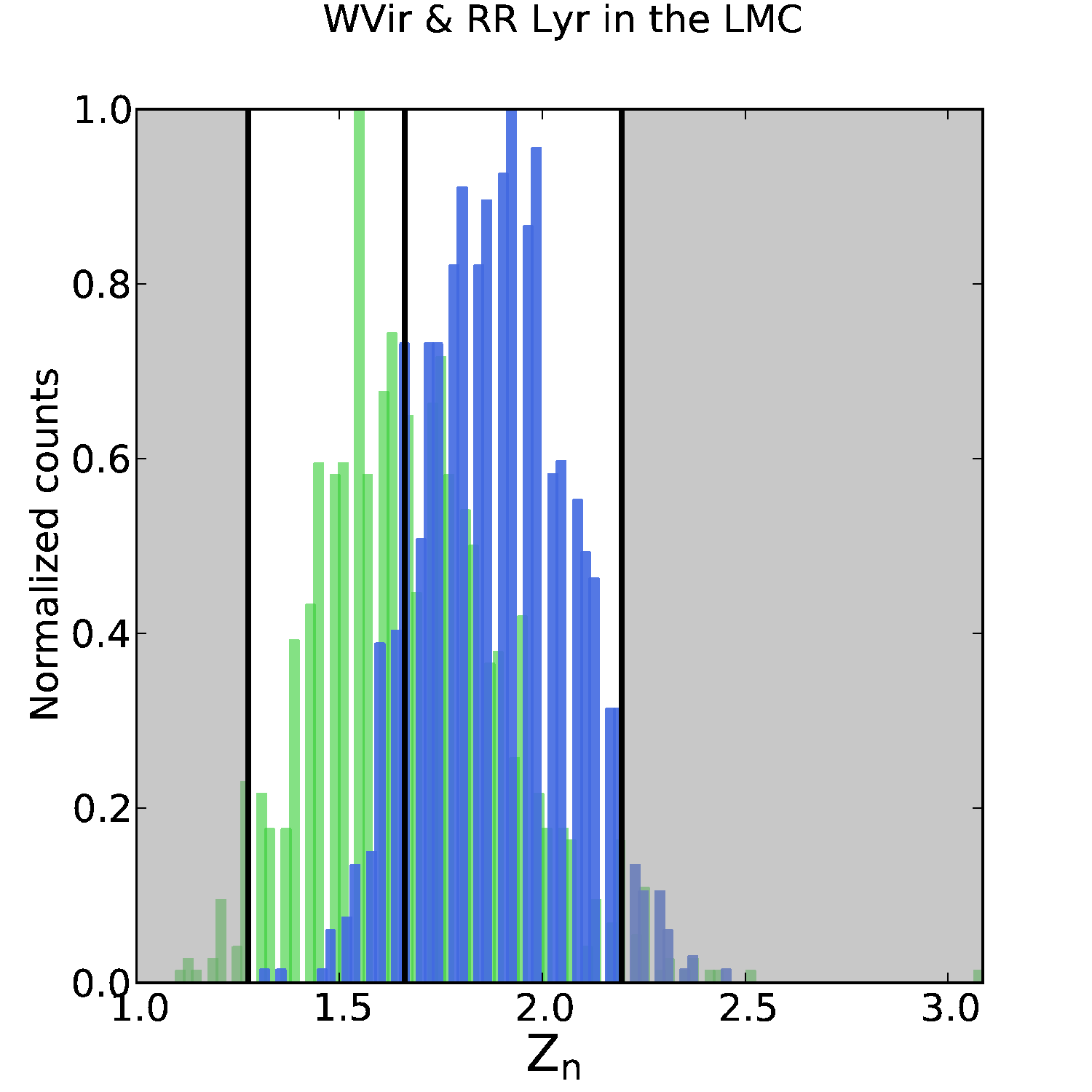}}
\FigCap{Distribution of the test statistic $Z_n$ for statistical test of
W~Vir stars with RR~Lyr variables after the cut of outskirts regions
in the LMC.}
\end{figure}

\subsection{Anomalous Cepheids}
The distribution of ACEPs in the Magellanic Clouds significantly differs
from the distribution of DCEPs -- all test statistics are in the rejection
area of the null hypothesis (Fig.~4e, LMC and SMC). In the tests of ACEPs
with RR~Lyr stars (Fig.~4f, LMC and SMC), some of the test statistics are
in the acceptability region. For the LMC, this value is 7.50\%, whereas in
the SMC it is 74.10\%. Therefore, in the SMC this value is close to our
acceptability criterion. Based on the statistical tests results, it is
impossible to draw unambiguous conclusions. However, looking on the spatial
distributions of ACEPs in comparison to RR~Lyr stars distributions
(Fig.~7), we can notice that ACEPs form a vast halo-like old population in
the LMC, and even larger halo than RR~Lyr variables in the SMC. Hence, it
seems that ACEPs belong to the old population.

\section{Comparison of the Decision Making Methods}
Gosset (1987) provided asymptotic equations for estimating the probability
$p$ and making the decision about the hypotheses. These formulae are defined
as follows:
$$1-\frac{Z_n}{Z_\infty}=0.75\cdot\left(\frac{n_1n_2}{n_1+n_2}\right)^{-0.9},\qquad
p\simeq2\exp{(-3(Z_\infty-1.05)^2)}.\eqno(11)$$ 
By using them it is possible to calculate asymptotic value of test statistic
$Z_\infty$ for sample sizes $n_1$ and $n_2$ for which statistical test gives
test statistic $Z_n$. We compared our results with the results obtained using
the Gosset's equations. As before, we assumed a significance level $\alpha=
0.05$. It means, that if $p\leq0.05$, we have to reject the null hypothesis,
but if $p>0.05$ there is no reason to reject the $H_0$, so we accept it. For
testing hypotheses using the Gosset's equations we decided to use
significantly larger samples of classical pulsators. In general, we used all
T2CEPs and ACEPs. Due to the very long computational time needed for
calculation the test statistics for large samples in the three-dimensional
space, we drew 20 times larger samples of stars from the LMC RR~Lyr and DCEPs
distributions. This means that we used 79 BL~Her stars and 1580 DCEPs or
RR~Lyr variables, 94 W~Vir stars and 1880 DCEPs or RR~Lyr variables, and we
tested 133 ACEPs with 2660 DCEPs or RR~Lyr stars. In the SMC, we used the same
number of DCEPs and RR~Lyr variables in every case.

In the LMC, we have 100\% agreement of both decision-making methods. In the
SMC, one out of the six statistical tests gives a different result. There
is a difference in the case of W~Vir and RR~Lyr stars test, where using the
Gosset's equations the null hypothesis should be accepted. However, our
result is very close to the acceptability criterion (the percentage of
$Z_n$ outside the rejection area is 83.50\%). We find that the Gosset's
equations do not work well in every case (\eg for tested pair BL~Her stars
with RR~Lyr variables, probability $p$ is greater than~1).

It should be noted that equations given by Gosset (1987) are only asymptotic
approximations, therefore it is much better to rely on the critical values of
the test statistic generated for the specific case. However, for simplified
cases it is enough to use the equations given by Gosset (1987). A single
different result out of all 12 tests means that the methods compatibility is
at a 92\% level.

\Section{Conclusions}
In this work, we compared three-dimensional distributions of various
classical pulsators. We used the OGLE collection of DCEPs, T2CEPs, ACEPs and
RR~Lyr stars (Soszyñski \etal 2015a,b, 2016, 2017, 2018), and PL
relations which we separately fitted to each group of pulsating stars. To
compare the spatial distributions, we used the extended Kolmogorov-Smirnov
test (Peacock 1983, Gosset 1987, Xiao 2017, {\it https://CRAN.R-project.org/package=Peacock.test}). As a main decision-making
method, we used mock ``theoretical'' distributions based on all DCEPs and
RR~Lyr stars, for which we counted critical values of the test statistics. We
compared our method with the asymptotic formulae given by Gosset (1987). We
omitted peculiar W~Vir and RV~Tau stars in the analysis, because the
distances to them were determined with large uncertainty.

Our results show that BL~Her stars have a similar spatial distribution to
that of RR~Lyr variables. They clearly form an extended halo similar to the
old population, providing the evidence that BL~Her stars are likely
old. Moreover, four of BL~Her stars from our sample are probable members of
the LMC globular clusters that are known to be old (Soszyñski \etal
2018). The masses of these variables are slightly smaller than those of
RR~Lyr stars (Groenewegen and Jurkovic 2017b). Additionally, modern
evolutionary scenarios successfully reproduce the evolution of these
stars. All the above conclusions allow us to infer, that BL~Her stars have an
age comparable to RR~Lyr variables.

The comparison of the spatial distribution of W~Vir stars to DCEPs and
RR~Lyrs (also with halo cut, which is described in Section~5) suggests that
the former variables are intermediate-age stars with age somewhere between
DCEPs and RR~Lyr stars. However, we found that one W~Vir variable is a
probable member of the LMC old globular cluster (Soszyñski \etal 2018).
For this reason, we are more inclined to conclude that these variables are a
mixture of old and intermediate-age stars. An additional argument for this
can be the masses of W~Vir stars -- they are comparable to BL~Her stars
masses (Groenewegen and Jurkovic 2017b). Moreover, following this we can also
state that they are comparable to RR~Lyr stars masses.

In the classical scenario of the T2CEPs evolution proposed by Gingold (1976,
1985), BL~Her and W~Vir variables should follow exactly the same spatial
distributions, because both groups belong to the same stellar population. Our
research shows, that distributions of considered T2CEPs subgroups are not
exactly the same, which questions the Gingold's scenario. The other piece of
evidence is that blue loops through the instability strip are not reproduced
in modern evolutionary calculations. This implies that we need other ideas
about evolution of W~Vir stars, which can be confirmed observationally.

Previous studies of the spatial distribution of ACEPs show that it is
different than distributions of DCEPs or RR~Lyr (Fiorentino and Monelli,
2012). Our results also do not provide unambiguous conclusions and are
consistent with the previous ones. For ACEPs in the Magellanic Clouds, we can
conclude with great certainty that three-dimensional distribution of these
stars is completely different than the distribution of DCEPs. The statistical
tests of the spatial distribution of these stars with RR~Lyr variables do not
give clear conclusions, because only in the SMC the majority of the test
statistics are outside the rejection area. However, looking at the results
from the SMC and having in mind the vast halo of ACEPs in both Clouds, we are
inclined to the theory, that these stars belong to the old population. Their
ages similar to the RR~Lyr variables suggest that the most favorable
evolution scenario for these stars is coalescence of two low-mass stars in a
binary system. If these stars evolved as a~single one, they should be
definitely younger.

The similarities of the three-dimensional distributions of T2CEPs and ACEPs
to the other classical pulsator distributions provide us with clues about the
origin of these stars and their properties. However, our research is based on
a statistical method, which by design requires numerous assumptions. In the
SMC, differences in the distributions of the pulsating stars, compared with
the LMC, are visible by the naked eye (\eg the lack of spiral arms and bar
in the DCEPs distribution). Therefore, we do not expect that the agreement
for both Magellanic Clouds will be at 100\%. Due to a greater number of
such stars in the LMC, future analyses will be more statistically meaningful
than the ones for the SMC.

\Acknow{We are grateful to the anonymous referee for suggestions and comments
  that greatly improved this publication. We would like to thank M. Marconi
  and M. Groenewegen for a fruitful discussion and valuable comments during
  4th MIAPP programme 2018 ``The Extragalactic Distance Scale in the Gaia
  Era'' which took place from  June 11 to July 6, 2018, in Garching,
  Germany. We also thank J.A. Peacock and Y. Xiao for advice regarding the
  idea of an multidimensional Kolmogorov-Smirnov test and its implementation
  in the R software.

  This work has been supported by the National Science Centre, Poland, grant
  MAESTRO no. 2016/22/A/ST9/00009 to I. Soszyñski. P. Iwanek is partially
  supported by the Kartezjusz programme no. POWR.03.02.00-00-I001/16-00
  founded by The National Centre for Research and Development, Poland. The
  OGLE project has received funding from the National Science Center, Poland,
  grant MAESTRO no. 2014/14/A/ST9/00121 to A. Udalski.}

\end{document}